%% file: plane.tex
\documentclass[12pt]{thesis}
\usepackage{algorithmic, algorithm, amsmath, amsthm, epsfig}

\newtheorem{lemma}{Lemma}[chapter]
\newtheorem{corollary}[lemma]{Corollary}
\newtheorem{theorem}[lemma]{Theorem}
\newtheorem{tree}{Tree Decomposition TD$\!\!$}[chapter]
\newtheorem{plane}{Plane Decomposition PD$\!\!$}[chapter]

\newcommand{\pw}[1]{$planewidth_{PD{#1}}$}
\newcommand{\tw}[1]{$treewidth_{TW{#1}}$}
\newcommand{\pd}[1]{PD{#1}$(G)$}
\newcommand{\td}[1]{TD{#1}$(G)$}
\newcommand{\lb}[1]{$lb({#1})$}
\newcommand{\Rb}[1]{$rb({#1})$}
\newcommand{\level}[1]{$level({#1})$}
\newcommand{\num}[1]{$num({#1})$}

\renewcommand{\maketitle}{
  \begin{titlepage}
    \begin{center}
	\vspace{.5in}
	\LARGE{Plane Decompositions as Tools for Approximation} \\
	\vspace{1in}
	\LARGE{Melanie J. Agnew}\\%\\mja3749@cs.rit.edu}\\
	\Large{mja3749@cs.rit.edu\\Rochester Institute of Technology\\Department of Computer Science\\102 Lomb Memorial Drive\\Rochester, NY 14623}\\
	\vspace{.5in}
    \LARGE{Christopher M.~Homan}\\
	\Large{http://www.cs.rit.edu/~cmh\\Rochester Institute of Technology\\Department of Computer Science\\102 Lomb Memorial Drive\\Rochester, NY 14623}\\
\vspace{.5in}
    \large{Date: \today}

    \end{center}
  \end{titlepage}
}

\begin{document}
\maketitle
\tableofcontents
\listoffigures 
\begin {abstract}
Tree decompositions were developed by Robertson and Seymour \cite{minor1}.  Since then algorithms have been developed to solve intractable problems efficiently for graphs of bounded treewidth.  In this paper we extend tree decompositions to allow cycles to exist in the decomposition graph; we call these new decompositions plane decompositions because we require that the decomposition graph be planar.  First, we give some background material about tree decompositions and an overview of algorithms both for decompositions and for approximations of planar graphs.  Then, we give our plane decomposition definition and an algorithm that uses this decomposition to approximate the size of the maximum independent set of the underlying graph in polynomial time. 
\end{abstract}
 
\chapter {Introduction} 
A graph decomposition is the grouping of the vertices of a graph $G$ into bags which are then connected to form a new graph.  A vertex of $G$ may appear in more than one bag and generally we require that each vertex appears in at least one bag.  For a decomposition $D = (S, P)$ of $G$, we call $V(P)$ the nodes of the decomposition graph $P$.  $S$ is the set of bags, one for each node of $P$.  \par
Graph decompositions are useful from both graph interpretation and calculation perspectives.  When drawing (and interpreting) a graph, they allow us to view the graph as if from a distance, observing relationships between groups of vertices.  This is useful because if the input set is large enough, a drawing of the graph can easily become cluttered with vertices and edges.  By grouping the vertices into bags and drawing only the relationships between the bags, we allow the user to gain a broader perspective of the graph before viewing the individual relationships present in each bag.  Algorithmically, decompositions allow us to find polynomial time algorithms for problems that would otherwise be computationally impossible for large graphs.  Given a graph with bounded decomposability we can use algorithms such as those given by Arnborg~\cite{arnborg1} to find maximum independent sets or maximum cliques efficiently.  \par
Tree decompositions specifically have been well studied since they were introduced by Roberson and Seymour \cite{minor1}.  Since then, Bodlaender has developed efficient algorithms both to find decompositions of graphs with bounded treewidth \cite{bodlaender96}, and to solve problems such as finding the maximum independent set of a graph given a tree decomposition \cite{bodlaender97}.  The tree decomposition definition is quite restrictive, but by requiring that the decomposition graph is a tree we are able to leverage powerful dynamic programming techniques for solving hard problems.\par
In developing plane decompositions, we were seeking a decomposition that is more intuitive from a viewing standpoint, realized by allowing cycles in the decomposition, while not sacrificing algorithmic power.  However, the application of algorithms for decompositions becomes less straightforward when we allow cycles in the decomposition.  (We can no longer start at the bottom of the tree and work upwards utilizing the separation between the top and bottom of the tree.)  There are relatively few algorithms for solving problems on planar graphs in polynomial time when compared to those for solving problems on trees.  We do have approximation algorithms however, one of which was developed by Baker \cite{baker}, who gives an approximation algorithm for planar graphs that can be used to solve problems that are generally (i.e. when restricted to just planar graphs) hard to approximate such as finding the maximum independent set. \par
In this paper we provide an investigation of plane decompositions (as an extension of tree decompositions).  When first developing a decomposition that allows for cycles we tried several straightforward modifications of the tree decomposition definition that proved to be less than ideal from a non-triviality perspective meaning that we ended up with decompositions that could easily be converted to tree decompositions or all graphs had a planewidth of 1.  As we continued, we shifted our focus towards the requirements of the dynamic programming algorithms, and more specifically to Baker's algorithm,  which allowed us to develop several more promising definitions.  Our final definition (Definition PD\ref{p7}) does not meet all of the criteria we give in chapter 6 but it does allow us to modify Baker's algorithm to use plane decompositions to approximate maximum independent sets of the underlying graph.  We've included several of our earlier definitions to provide the reader with some insight into the tension between what seem like ideal properties of a plane decomposition (for example planar graphs should have a planewidth of 1) and algorithmic utility. \par
We begin with background material in chapter 1 and we give a brief discussion of tree decompositions in chapter 2.  In chapter 3 we provide an overview of decomposition algorithms in general and in chapter 4 we give a description of Baker's algorithm for approximating maximum independent sets  of planar graphs.  We give these overviews early in the paper because they were influential in shaping our final plane decomposition definition.  Chapter 5 is a progression of plane decomposition definitions with a narrative of their evolution.  In chapter 8 we provide a discussion of how Baker's algorithm can be modified to approximate independent sets of a graph given a plane decomposition that fits Definition PD\ref{p7} and in chapter 7 we conclude with a complexity analysis of our algorithm and possible ideas for further research.  Although the work described here is theoretical we provide an implementation of our algorithm in appendix A.

\chapter{Background}
We begin with a series of definitions related to graphs, a more lengthy discussion can be found in a textbook such as the one by West~\cite{west}.  Formally, a graph $G$ contains a set of vertices, $V(G)$, and a set of edges which connect the vertices, $E(G)$.   If an edge $e$ connects two vertices $u$ and $v$, then $e$ can be written as $uv$ or $vu$. The number of vertices in $G$ is denoted by $||V(G)||$.  For $v \in V(G)$, the neighborhood of $v$, denoted by $\Gamma(v)$, is the set $\{u \mid uv \in E(G)\}$.  For $F \subseteq V(G)$, $\Gamma(F) = \{u \mid v \in F \wedge u \in \Gamma(v) \wedge u \notin F\}$.  We assume for the purposes of this paper that graphs are simple, meaning that edges are unweighted and undirected and that there are no self loops (i.e. ($\forall~ v \in V(G))$ $[vv \not\in E(G)]$). \par
$G'$ is a subgraph of $G$ if and only if $V(G') \subseteq V(G)$ and $E(G') \subseteq E(G)$.  We also call $G$ a supergraph of $G'$.  If $G'$ is a subgraph of $G$ induced by $F \subseteq V(G)$, then $G'$ has vertex set $F$ and edge set $\{uv \mid uv \in E(F)\}$.  We write $G \; \backslash \; F$ to mean the subgraph induced by $V(G) - F$.\par
A path is a graph $G$ with $V(G) = \{v_{1}$, $v_{2}$, ..., $v_{n}\}$ and $E(G) = \{v_{i}v_{i+1} \mid 1 \leq i < n\}$.  A cycle is a path with one extra edge connecting vertices $v_{n}$ and $v_{1}$.  A chord of cycle $C$ is an edge not in $C$ whose endpoints lie in $C$.  A graph is triangulated if it contains no chordless cycles of length $> 3$.  A tree is a connected graph with no cycles.  A clique $K$ is a graph such that $(\forall~ u$, $v \in V(K): u \not= v)$ $[uv \in E(K)]$.  We use $K_{i}$ to denote a clique of $i$ vertices.  An independent set $I$ is a graph such that $(\forall~ u$, $v \in V(I): u \not= v)$ $[uv \notin E(I)]$.\par
A graph $G$ is connected if and only if for every $u, v \in V(G)$ there is a path in $G$ between $u$ and $v$.  The components of $G$ are its maximal connected subgraphs.  A separator of $G$ is a set $S \subseteq V(G)$ such that $G \; \backslash \; S$ has more than one component. $G$ is $k$-connected if $k$ is the minimum size of a set $S \subseteq V(G)$ such that $G \; \backslash \; S$ is disconnected.\par
Graph drawing is the field of study centered around the representation of graphs, often in a two-dimensional plane.  Vertices are commonly drawn as symbols such as circles or squares while edges are represented by simple, finite curves.  More formally, a drawing of a graph $G$ is a pair of mappings $f$ and $f'$.  For each $v \in V(G)$, $f(v)$ maps $v$ to a point in the plane; for each $e \in E(G)$, $f'(e)$ maps $e$ to a simple, closed curve in the plane with endpoints $u$ and $v$.  For all $u$, $v \in V(G)$, $f(u) \not= f(v)$ and for each $e$, $e' \in E(G)$, $f'(e) \cap f'(e')$ contains at most a single point.  If for edges $e$, $e' \in E(G)$ there is a point $p$ in $f'(e) \cap f'(e')$ that is not a shared endpoint of $e$ and $e'$, $p$ is an edge crossing.\par
A drawing of graph $G$ that does not contain any crossings is called a planar embedding of $G$.  $G$ is considered to be planar if there exists a planar embedding of $G$. Figure \ref{k4} shows three drawings of $K_{4}$ (a clique with four vertices); \ref{k4}.a has one edge crossing, and \ref{k4}.b and \ref{k4}.c are planar embeddings.  If $G$ is planar, then there exists a planar embedding of $G$ using only straight lines for edges.  Figure \ref{k4}.c is a straight line embedding. \par
\begin{figure}[htb] \begin{center}  \input{k4.pstex_t}\\ Embeddings b and c are planar, both are also 2-outerplanar. \caption{Three drawings of $K_{4}$.} \label{k4}\end{center}\end{figure}
The faces of a planar embedding of a graph are the maximal connected regions of the plane that do not contain any vertices or edges used in the embedding (i.e. the areas enclosed by the cycles of the graph).  In figures \ref{k4}.b and \ref{k4}.c there are four faces.  The exterior face of an embedding is the unbounded face, the other faces are called interior faces.  Harary defines a term outerplanar; an embedding $f$ is outerplanar if for all $v \in V(G)$, $v$ is on the exterior face of $f$ \cite{harary}.  Baker~\cite{baker} extends this idea by introducing levels and $k$-outerplanar graphs.  For graph $G$ and embedding $f$ of $G$, vertex $v \in V(G)$ has $level(v) = 1$ if $v$ is on the exterior face of $f$.  Vertex $v$ has \level{v} $= l$ if $v$ is on the exterior face of $f$ with the vertices $\{u \mid level(u) < l\}$ removed, meaning $v$ is on the outer face of $G  \; \backslash \; \{u \mid level(u) < l\}$.  Embedding $f$ is $k$-outerplanar if there does not exist $v \in V(G)$ such that \level{v} $> k$.  The embeddings in figures \ref{k4}.b and \ref{k4}.c are 2-outerplanar with \level{a} = \level{b} = \level{c} $= 1$ and \level{d} $= 2$.  The embedding given in figure \ref{baker2} is 3-outerplanar, the vertices on the outer face are labeled 1:$x$ and are level 1 vertices; the level 2 and 3 vertices are labeled 2:$x$ and 3:$x$ respectively.  \par
It is interesting to note that although not all graphs are planar, if we extend our embeddings to three dimensions, then all graphs can be drawn with straight line edges and without crossings.  For example, if for each vertex $v_{i} \in V(G)$ we map $v_{i}$ to $(i$, $i^2$, $i^3)$, then we have an embedding on a three dimensional grid with no edge crossings.  Further discussion of this and similar constructions can be found in a paper by Dujmovi\'c, Mworin, and Wood~\cite {dmw}.\par
\begin{figure}[htb] \begin{center}  \input{c6.pstex_t} \caption{Two drawings of $C_{6}$.} \label{c6}\end{center}\end{figure}
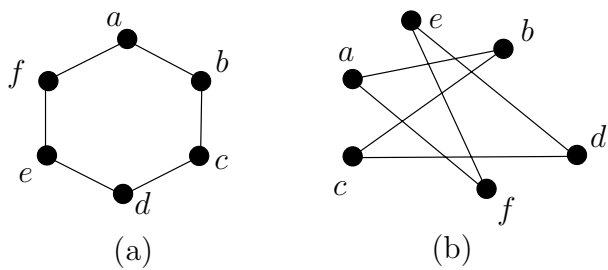

Often our goal in choosing a particular drawing with which to represent a graph is to maximize readability.  Although it is not clear that one particular drawing property is key in affecting readability, minimizing the number of crossings does seem to have a positive effect.  Consider, for example, the two embeddings of $C_{6}$ (a cycle with six vertices) shown in figure \ref{c6}.  With just a quick glance at figure \ref{c6}.a we are able to determine the relationships between the vertices.\par  
Other drawing properties we might consider are the number of bends in the edges, overall symmetry, the angles between the edges leaving an individual vertex, and the angles at which edges intersect.  Eades and Tamassia~\cite{eades88} give an overview of graph drawing algorithms including a discussion of properties of drawings.  Purchase~\cite{purchase} studied the effects of optimizing the properties given above (i.e. minimizing edge bends, minimizing crossings, maximizing symmetry, etc) in performing such tasks as finding the shortest path between two vertices and found that minimizing edge crossings gave slightly better readability results.  \par
Planar graphs offer benefits from an interpretation viewpoint, and in some cases provide an algorithmic advantage.  Before we give examples of problems for which planar graphs offer an advantage, we give a brief discussion of problem complexity.  There are many classes we can use to identify graph problems in terms of their hardness.  We give a brief overview of several of the main classes here.  Problems that can be solved in polynomial time are in the class P.  Problems with solutions that can be verified in polynomial time by a nondeterministic Turing machine are in the class NP.  The class P is contained in the class NP, and NP is thought to contain problems which are not in the class P.  If P and NP are not equivalent, then the solution for an NP problem may, in the worst case, require an exhaustive search.  A problem is NP-hard if an algorithm giving a solution for it can be used to find a solution for any other problem in NP.  A problem is NP-complete if it is NP-hard and can be verified in polynomial time by a nondeterministic Turing machine.  There are also problems that are hard even to approximate, in these cases we refer to an approximation ratio $\alpha \geq 1$, where solutions are required to be at least $OPT/\alpha$ (for maximization problems) where $OPT$ is some measure of the optimal solution. A problem $A$ is NP-hard to approximate if we can reduce solutions to that problem to within a factor $\alpha$ of another NP-hard problem $B$.  The ratio of the solutions for $B$ to the solutions for $A$ (assuming maximization problems) is then $OPT/\alpha$. For further discussion of complexity classes, see for example the text by Hopcroft, Motwani and Ullman \cite{hopcroftb}.)\par
Returning to our discussion of specific problems and planar graphs, the coloring problem (assigning each vertex a color such that no adjacent vertices have the same color, using the minimum number of colors) can be solved in polynomial time (for four or more colors) when limited to only planar graphs \cite{roberts96}.  Max Cut (also an NP-hard problem on general graphs) can be solved in polynomial time when restricted to planar graphs \cite{hadlock}.  We can also test to see if a graph is planar, and produce a planar embedding if one exists, in linear time using the algorithm by Hopcroft and Tarjan~\cite{hopcroft74}.  We can modify non-planar graphs by removing edges or adding vertices to make the resulting graph planar, Liebers~\cite{liebers} gives a survey of  these techniques.\par
There are other problems such as maximum independent set (given a graph $G$ and integer $k$, does $G$ contain an independent set $I$ with $||V(I)|| \geq k$) that are NP-complete for planar graphs (and are hard even to approximate on general graphs) but have polynomial time approximation algorithms for planar graphs \cite{baker, chiba, lipton}.  Baker \cite{baker} presents a method that can be used to approximate solutions on planar graphs for many NP-complete problems such as maximum independent set and minimum vertex cover.  (A vertex cover of a graph $G$ is a set $F \subseteq V(G)$ such that $(\forall~ uv \in E(G))$, $[u$ or $v$ is in $F]$.)  See Garey and Johnson~\cite{garey} for a formal definition of this and the other NP-complete problems named in this paragraph.  If graph $G$ is $k$-outerplanar for a fixed $k$, we can use Baker's algorithm to solve the above problems in polynomial time, as well as Hamiltonian circuit and Hamiltonian path.  (A Hamiltonian path is a path in $G$ such that each vertex is visited exactly once.)  We will use the Baker algorithm to show that we can approximate a maximum independent set of a graph $G$ with fixed planewidth in polynomial time, and find a maximum independent set of $G$ with a plane decomposition $D$ that is a $k$-outerplanar graph for fixed $k$.

%%%%%%%TREE%%%%%%%%%%%%%%%%%%%%%%%%
\chapter{Tree Decompositions}
The notion of a tree decomposition was developed by Robertson and Seymour in their study of graph minors~\cite{minor1, minor2}.  Since then, many algorithms have been developed both to find tree decompositions of different types of graphs~\cite{bodlaender96} and to use tree decompositions to solve intractable problems.  Although the problem of determining if a given graph has a treewidth $\leq k$ is NP-complete, for any fixed $k$, linear time algorithms exist which, given a graph $G$, output a tree decomposition with treewidth $\leq k$ or determine that no such decomposition exists~\cite{bodlaender97}.  \par
%%%%%%%%%%%%tree decomposition
\begin{tree}\label{t1}\cite{minor1} Given a graph $G $, a tree decomposition of the graph is a pair $(S$, $T)$ with $S = \{X_{i} \mid i \in V(T)\}$ a family of subsets of $V(G)$, one for each node of $T$, such that 
\begin{enumerate}
\item $\bigcup_{i \in V(T)} X_{i} = V(G)$,
\item $(\forall~ uv \in E(G))$ $(\exists~ \; i \in V(T))$ $[u$, $v \in X_{i}]$, and 
\item $(\forall~ i$, $j$, $k \in V(T))$ $[(j$ is on the path from $i$ to $k$ in $T) \Rightarrow (X_{i} \cap X_{k} \subseteq X_{j})]$.
\end{enumerate} 
The width of a tree decomposition is $max_{i \in V(T)} ||X_{i}|| - 1$.  The \tw{\ref{t1}} of a graph $G$ is the minimum width over all tree decompositions of $G$.\end{tree} 
For example, graph $G$ is drawn in figure \ref{g1} and a decomposition $D \in $ \td{\ref{t1}} with width 3 is given in figure \ref{d1}.  For each node $i$, we have listed the contents of $X_{i}$ inside the circle for $i$.  This is not the only tree decomposition of $G$.  For instance, we could have one node containing all the vertices of $G$, or we could split node 1 into two nodes $i$ and $j$ with $X_{i} = \{a$, $b$, $c\}$ and $X_{j} = \{b$, $c$, $d\}$.  Then nodes 2 and 3 would be children of $i$ and node 4 would be a child of $j$.  We could also move node 3 to be a child of node 2.  If we were to make node 2 a child of node 5, we would need to add vertex $a$ to $X_{5}$.  Note that $G$ has a clique of size 4 (the set $\{a$, $c$, $d$, $e\}$) and the \tw{\ref{t1}} of $G$ is 3.\par
\begin{figure}[htb] \begin{center}  \input{g1.pstex_t} \caption{A graph $G$ with \tw{\ref{t1}} of 3.} \label{g1}\end{center}\end{figure}
\begin{figure}[htb] \begin{center}  \input{d1.pstex_t}\\The vertices of $G$ are labeled inside the nodes, the numbers on the outside of the nodes denotes their number in $V(P)$. \caption{Decomposition $D \in$ \td{\ref{t1}} of graph $G$ given in figure \ref{g1}.} \label{d1} \end{center}\end{figure}
An alternate definition for treewidth exists that takes a more constructive approach.  
\begin{tree} \label{t2}\cite{kloks} A clique with $k+1$ vertices is a $k$-tree. Given a $k$-tree $T_{n}$ with $n$ vertices, a $k$-tree with $n+1$ vertices is constructed by adding a new vertex $x_{n+1}$ to $T_{n}$; $x_{n+1}$ is made adjacent to a $k$-clique of $T_{n}$ and nonadjacent to the other $n-k$ vertices of $T_{n}$.\end{tree}
The following lemmas give us some characteristics of tree decompositions which will be useful in evaluating our proposed plane decomposition definitions. 
Cliques are in some sense the ``least planar'' graphs since they contain all possible edges.  In chapter 6 we discuss the planewidth of cliques for our various plane decomposition definitions and so it is useful to have a similar result for the treewidth of cliques to use for comparison.  It appears from reading relevant literature that it is a well known fact that the treewidth of a clique of $n$ vertices is $n-1$ but we were unable to find a proof so we give this result with the following lemma and corollary.

%%%%%%%%%%%%%%%%%%%%%%% there is a bag in the tree decomp containing clique k 
\begin{lemma}\label{treeclique} For graph $G$ and decomposition $D \in$ \td{\ref{t1}} with $D = (S,T)$, if $G$ is a clique, then there is a node $i \in V(T)$ with $V(G) \subseteq X_{i}$.\end{lemma}
\begin{proof}[Proof by induction]
Let $G$ be a graph and $D = (S$, $T)$ a decomposition such that $D \in$ \td{\ref{t1}}.  Suppose $||V(T)||=1$, then from Definition TD\ref{t1}.1, there exists an $i \in V(T)$ such that $V(G) \subseteq X_{i}$.  Suppose $||V(T)|| = 2$, with $V(T) = \{1$, $2\}$.  It follows immediately from Definition TD\ref{t1}.2 that either $V(G) \subseteq X_{1}$ or $V(G) \subseteq X_{2}$.  Assume that for $||V(T)|| = n$ with $1 \leq n \leq ||V(T)||$, there exists an $i \in V(T)$ with $V(G) \subseteq X_{i}$.\\
Induction hypothesis: For $||V(T)|| = n+1$ and $2 \leq n < V(T)$,  there exists an $i \in V(T)$ with $V(G) \subseteq X_{i}$.  Let $a$ be leaf node in $V(T)$, and let $b$ be $a$'s parent in $T$.  Either $V(G) \subseteq X_{a}$, or from Definition TD\ref{t1}.2, for all $v \in X_{a}$, there exists an $i \in V(T)$ with $a \not= i$ such that $v \in X_{i}$.  If the former is the case, clearly there exists an $i \in V(T)$ with $V(G) \subseteq X_{i}$.  If the latter is the case, from Definition TD\ref{t1}.3, $X_{a} \subseteq X_{b}$.  Consider decomposition $D' = (S', T')$ with $V(T') = V(T) - \{a\}$, $E(T') = E(T) - \{ab \}$, and $S'$ = $S - \{X_{a}\}$ In other words, $T'$ tree $T$ with node $a$ removed.  $D'$ is a tree decomposition of $T$ with $n$ nodes, therefore there exists an $i \in V(T')$ with $V(G) \subseteq X_{i}$.
\end{proof}

\begin{corollary} For a clique $G$ with $||V(G)|| = n$, the \tw{\ref{t1}} of $G$ is $n-1$. \end{corollary} 
\begin{proof} If $G$ is a clique with $n$ vertices, then from Lemma \ref{treeclique} we know that decomposition $D \in$ \td{\ref{t1}} contains a bag $i$ such that $i \in V(T)$ and $V(G) \in X_{i}$.  Thus the width of $D$ is $n-1$ from Definition \ref{t1} and the \tw{\ref{t1}} of $G$ is also $n-1$.  \end{proof}

The following lemma describes the relationship between separators in the tree decomposition and separators in the graph.  This property is useful when we are developing efficient algorithms for decompositions.
\begin{lemma}\label{sep1} For graph $G$ and tree decomposition $D = (S, T)$, if $J \subseteq V(T)$ is a separator of $T$ and $C_{1}, ..., C_{n}$ are the connected components of $T \; \backslash \; J$, then $(\forall~ i$, $j$: $i \not= j)$  $(\forall~ x \in \bigcup_{k \in C_{i}}X_{k})$ $(\forall~ y \in \bigcup_{k \in C_{j}}X_{k})$ $[(\{x,y\} \cap \bigcup_{k \in J} X_{k} = \emptyset) \Rightarrow (xy \notin E(G))]$.\end{lemma}
\begin{proof} Assume that $(\exists~ i$, $j$: $i \not= j)$  $(\exists~ x \in \bigcup_{k \in C_{i}}X_{k})$ $(\exists~ y \in \bigcup_{k \in C_{j}}X_{k})$  such that $((\{x,y\} \cap \bigcup_{k \in J} X_{k} = \emptyset)$ $\wedge$ $(xy \in E(G)))$.  See figure \ref{tree_sep}.  By Definition TD\ref{t1}.2, there exists a $p \in V(T)$ st $x, y \in X_{p}$.  Assume WLOG $p \notin C_{j}$ (since one node cannot appear in two components).  By Definition TD\ref{t1}.3, if $x \in \bigcup_{k \in C_{i}}X_{k}$ and $x \in X_{p}$, $x$ must also be in every node on the path between $i$ and $p$.  Since $p \notin C_{i}$, there exists a $q$ on the path between the node(s) in component $C_{i}$ containing $x$ and $p$ that is also in $J$ such that $x \in X_{q}$, this is a contradiction.  
\begin{figure}[htb] \begin{center}  \input{tree_sep.pstex_t}\\For tree decomposition $D = (S, T)$, vertex $x$ must be in $X_{q}$ for some $q \in J$,  because $q$ is on the path between two components which contain $x$. \caption{Drawing for Lemma \ref{sep1}} \label{tree_sep} \end{center}\end{figure}\end{proof}

Intuitively speaking, this means that if two vertices, $x$ and $y$ can be separated in the decomposition of $G$, then they cannot be neighbors in $G$.  Referring back to figure \ref{d1}, if $J = \{1\}$ we have three components in $P \; \backslash \; J$; $C_{1} = \{2, 5, 6\}$, $C_{2} = \{3\}$, and $C_{3} = \{4, 7\}$.  Notice that for vertices $j$ and $f$, $j \in X_{3}$, $f \in X_{5}$ and $\{j$, $f\} \cap  X_{1} = \emptyset$.  Also, $3 \in C_{1}$ and $5 \in C_{2}$ and $jf \notin E(G)$. Vertex $d$ has neighbors in components $C_{1}$, $C_{2}$, and $C_{3}$; note that $d$ is therefore in $X_{1}$, $X_{2}$, $X_{4}$, and $X_{6}$.  For a graph $G$ and decomposition $D = (S, T)$ for $D \in$ \td{\ref{t1}}, because all of the non leaf bags in $T$ are separators of $T$, the non leaf bags of $T$ must contain separators of $G$.  Thus $G \; \backslash \; X_{i}$, for some $i \in V(T)$ such that $i$ is not a leaf node, should contain two separate components (unless one of the components of $T \; \backslash \; \{i\}$ is contained in $X_{i}$). \par

%%%%%%%%%%decomposition algorithms%%%%%%%%%%%%%%%%%%%%%%%%%
\chapter{Efficient algorithms using decompositions}
Bodlaender~\cite{bodlaender97} and Arnborg~\cite{arnborg1, arnborg2} give similar approaches for computing solutions to hard problems such as finding the Max IS of graphs with bounded treewidth.  These methods are based on a dynamic programming approach: tables are computed in a bottom up fashion,  for disjoint sections of the graph which are then combined so that a solution for the entire graph is found using only tables bounded in size by a factor of the treewidth.  Bodlaender~\cite{bodlaender97} starts at the bottom of the tree decomposition and computes one table for each node of the tree; the table for the root of the tree contains the size of the maximum independent set (Max IS) of the underlying graph.  Table $i$ contains a maximum of $2^{||X_{i}||}$ rows, each row consists of $s \subseteq X_{i}$, and a value  $is$ which is the size of the Max IS of the vertices in $\{v \mid v \in$ row $j\} \cup \{v \mid v \in X_k$ where $k$ is a node in the subtree with root $i$ and $v \notin X_{i}\}$ which includes the vertices in $s$.  We use the notation $t.r.c$ to denote the entry in table $t$, which corresponds to node $t$, with row $r$ and column $c$.  \par
For example, several tables for the decomposition shown in figure \ref{d1} are given in figure \ref{d1tables}.  We have not listed rows which contain invalid entries, such as the set $\{a$, $c\}$ for table 2 ($a$ and $c$ cannot both be in the independent set since $ac \in E(G)$).
\begin{figure}\begin{center}
\begin{tabular}{|l|c|l|} \hline
row & $s$ & $is$ \\ \hline
1 & $\emptyset$ & 0 \\ \hline
2 & $\{c\}$ & 1 \\ \hline
3 & $\{e\}$ & 1 \\ \hline
4 & $\{f\}$ & 1 \\ \hline
5 & $\{e$, $f\}$ & 2 \\ \hline
\end{tabular}\\ \vspace{.15in}  table for node 5\\\vspace{.3in}
\begin{tabular}{|l|c|l|}\hline
row & $s$ & $is$ \\ \hline
1 & $\emptyset$ & 0 \\ \hline
2 & $\{d\}$ & 1 \\ \hline
3 & $\{e\}$ & 1 \\ \hline
4 & $\{g\}$ & 1 \\ \hline
\end{tabular}\\ \vspace{.15in} table for node 6\\ \vspace{.3in}
\begin{tabular}{|l|c|l|}\hline
row & $s$ & $is$ \\ \hline
1 & $\emptyset$ & 2 \\ \hline
2 & $\{a\}$ & 2 \\ \hline
3 & $\{c\}$ & 2 \\ \hline
4 & $\{d\}$ & 2 \\ \hline
5 & $\{e\}$ & 2 \\ \hline
\end{tabular}\\ \vspace{.15in} table for node 2\\ \vspace{.3in}
\begin{tabular}{|l|c|l|} \hline
row & $s$ & $is$ \\ \hline
1 & $\emptyset$ & 4 \\ \hline
2 & $\{a\}$ & 3 \\ \hline
3 & $\{b\}$ & 3 \\ \hline
4 & $\{c\}$ & 3 \\ \hline
5 & $\{d\}$ & 3 \\ \hline
6 & $\{a$, $b\}$ & 2 \\ \hline
\end{tabular}\\ \vspace{.15in}  table for node 1\\
Four tables for the computation of the independent set of the graph given in figure \ref{g1} using the decomposition given in figure \ref{d1}
\caption{Tables for independent set computation.}
\label{d1tables}\end{center} \end{figure}
Table 5 gives us the sizes of the Max IS of the vertices $\{c$, $f$, $e\}$.  The set $5.1.s$ is empty, and $5.1.is = 0$ because the size of the Max IS of the set $\{c$, $f$, $e\}$ where none of these vertices is included, is zero.  $5.2.is = 1$ because we are including the vertex $c$ in the Max IS.  In row 5.5, $5.5.s = \{e$, $f\}$, since $ef \notin E(G)$, $5.5.is = 2$.\par
As the computation moves up the tree, only combinations of the vertices in $X_{t}$ appear in table $t$.  For example, table 2 gives us the size of the Max IS of the set $\{a$, $c$, $d$, $e$, $f$, $g\}$ based on the inclusion of the vertices in $X_{2}$ in the independent set.  For each group of vertices $2.i.s$ we look for rows in table 5 where the vertices in $X_{2} \cap X_{5}$ are included in the $s$ entries iff they are included in $2.i.s$.  We look for rows in table 6 where the vertices in $X_{2} \cap X_{6}$ are included in the $s$ entries iff they are included in entry $2.i.s$.  Then we maximize the sum $5.j.is + 6.k.is - ||5.j.s \cap 6.k.s||$ over rows $5.j$ and $6.k$ where the overlap of $X_{5}$ and $X_{6}$ match.  More formally we look for the rows that meet the following criteria:
\begin{enumerate}
\item $(2.i.s \cap X_{5}) = (2.i.s \cap 5.j.s)$, 
\item $(2.i.s \cap X_{6}) = (2.i.s \cap 6.k.s)$, and 
\item $5.j.s - (2.i.s \cap X_{5}) = 6.k.s - (2.i.s \cap X_{6})$
\end{enumerate}
For example, $2.1.is = 5.4.is + 6.4.is - 0$, meaning that the Max IS computed so far contains the vertices $f$ and $g$ but not $a$, $c$, $d$, or $e$.  $2.5.is = 5.5.is + 6.3.is - 1$, meaning that the Max IS computed so far contains the vertices $e$ and $f$ but not $a$, $c$, or $d$($e$ is included in both  $5.5.s$ and $6.3.s$ which is why we subtracted 1).  Computation continues in this manner until we reach the top of the tree.  By looking at table 1 we can see that the size of the Max IS for this graph is 4.  \par

%%%%%%%%%%%%%%%%%%%%BAKER%%%%%%%%%%%%%%%%%%
\chapter{Planar Graphs: Baker's Algorithm}
We now shift our attention from trees to planar graphs.  Since our goal is to develop a plane decomposition with an efficient algorithm, we now describe Baker's algorithm \cite{baker} which is an approximation algorithm for planar graphs.  Baker's algorithm uses separators of planar graphs to find an approximation of the maximum independent set (Max IS) in polynomial time.  This algorithm works by using dynamic programming to find the maximum independent set for a series of subgraphs of $G$, $G_{0}, ..., G_{k-1}$ where $G_{i}$ has the levels $\{l \mid i \equiv l \bmod(k+1)\}$ removed for some chosen positive integer $k$.  So, graph $G_{i}$ is the graph $G \; \backslash \; \{v \mid$ \level{v} $\bmod(k+1)  \equiv i \}$.  In figure \ref{bakersubs} we give two subgraphs of a graph $G$ for $k = 2$ (where levels with dashed lines are removed).  Notice that both of the given subgraphs are made up of disconnected 2-outerplanar graphs.
\begin{figure}[htb] \begin{center}  \input{baker_subs.pstex_t}\\ $G_{0}$ and $G_{2}$ of a graph $G$ for $k = 2$ used in Baker's algorithm, levels with dashed lines are removed. \caption{Two subgraphs of a graph $G$ for Baker's algorithm.} \label{bakersubs}\end{center}\end{figure}
By the pigeonhole principle, one of these graphs has a maximum of $1/(k+1)$ vertices in the independent set, so by calculating the Max IS for each of them, we can find an independent set that is $k/(k+1)$ optimal for $G$. This leads us to Baker's main theorems:

\begin{theorem}\cite{baker}\label{bakerth1}Let $k$ be a positive integer.  Given a $k$-outerplanar embedding of a $k$-outerplanar graph, an optimal solution for maximum independent set can be obtained in time $O(8^{k}n)$, where n is the number of vertices.\end{theorem}

\begin{theorem} \cite{baker}\label{bakerth2}For fixed $k$, there is an $O(8^{k}kn)$-time algorithm for maximum independent set that achieves a solution of size at least $k/(k+1)$ optimal for general planar graphs.  Choosing $k = \lceil c~log~log~n \rceil$, where $c$ is a constant, yields an approximation algorithm that runs in time $O(n(\log n)^{3c}log~log~n)$ and achieves a solution of size at least $\lceil c~log~log~n\rceil / (1 + \lceil c~log~log~n\rceil)$ optimal.  In each case, n is the number of vertices.\end{theorem}

We will first give an overview of Baker's algorithm, then in chapter 7 we will show how it can be modified to work with plane decompositions.  \par 
For each $G_{i}$ we have a series of $k$-outerplanar graphs for which we find the MAX IS independently (these values are then summed and used to find the Max IS for $G$).  For the rest of the description we focus our attention on a single $k$-outerplanar graph $G$.  We begin by constructing a $k$-outerplanar embedding $f$ of $G$ which can be done in polynomial time using an algorithm by Bienstock\cite{Bienstock}.  Next, we construct graph $G'$ from $G$ with $V(G') = V(G)$ and $E(G') = E(G) \bigcup F$ where $F \subseteq \{uv \mid $ \level{u}$-1 =$ \level{v}$\}$  is a set of edges that makes $G'$ triangulated while preserving it's planarity.  $f'$ is an embedding of $G'$ constructed by adding the edges in $F$ to $f$. Figure \ref{baker1} contains a planar embedding of a graph $G$. Figure \ref{baker2} gives a triangulated supergraph of $G$ where each vertex is labeled \level{v}:\num{v}, the edges in $F$ are dashed.  (We discuss how \num{v} is calculated below.)\par
Next, we divide $G'$ into slices.  A slice is a subgraph of $G'$ with two boundaries (similar to a slice of a pie).  The edges of a slice are paths in $G'$ which extend from a center component of $f'$ to the outermost face of $f'$.  If $G'$ has $n$ slices, $\{1, ..., n\}$, and $j \in \{1, ..., n\}$ is a slice of $G'$, then $\bigcup_{1\leq j \leq n}V(j) = V(G)$.  We call the boundaries of slice $j$ the left bound and right bound which we denote \lb{j} and \Rb{j} respectively (from Baker's notation).  For slice $j$, if $l = \max\{$\level{v} $\mid v \in V(j)\}$, then both boundaries of slice $j$ contain exactly one vertex from each level $\leq l$ which form a path in $G'$.  Thus, the total boundary of $j$ contains a maximum of $2l$ vertices.  As we travel around the graph in a counterclockwise fashion, slices $j$ and $j+1$ have \Rb{j} = \lb{j+1}.  Because $G'$ is planar, the nodes on the interior of slice $j$ (i.e. $V(j)~ - $ \lb{j} $-$ \Rb{j}) cannot be connected to the nodes on the interior of another slice $k$ for $j \not=k$.  \par
For each slice, we use dynamic programming to compute a table that contains the size of the largest independent set based on the inclusion of the boundary vertices.  When two slices, $j$ and $k$ for $j < k$ are merged, only vertices on their shared boundary can be connected to vertices interior to the slices.  Thus, we need only include in our new table combinations of \lb{j} and \Rb{k} that do not include any edges of $G$.  For the set $S \subseteq$ \lb{j} $\cup$ \Rb{k}, we find the set of vertices in the shared boundary of $j$ and $k$  that gives us the Max IS value for the set of nodes of the new boundary.\par
Before describing how the slices are constructed we give a description for how to label the vertices.  In her algorithm, Baker constructs trees and uses them to construct the slices; the method given here is an overview of how the algorithm works and is meant to facilitate a discussion of how this method applies to plane decompositions.  We assume $G$ is a connected graph.  If it were not, we would run the algorithm for the connected components of $G$ separately and sum their Max IS sizes to get a maximum for $G$.\par
\begin{figure}[htb] \begin{center}  \input{baker1.pstex_t} \caption{Graph $G$} \label{baker1}\end{center}\end{figure}
\begin{figure}[htb] \begin{center}  \input{baker2.pstex_t}\\Graph $G'$ is a triangulated supergraph of graph $G$ shown in figure \ref{baker1}.  The vertices are labeled \level{v}:\num{v} \caption{Triangulated graph $G'$.} \label{baker2}\end{center}\end{figure}
For a vertex $v$, we use \num{v} to denote its number; we use the term \emph{minimum vertex} to mean $\min\{$ \num{v} $\mid v \in U\}$ for some $U \subseteq V(G')$.  For example, we may refer to the minimum vertex of a particular component.  We use the methods \emph{Setup}, \emph{NumberGraph}, and \emph{NumberComponent} to number the vertices of $G'$.   First, for all $v \in V(G')$, \emph{Setup} initializes each $num(v)$ to zero.  \emph{NumberGraph} calls \emph{NumberComponent} once for each component of the graph after first determining the starting vertex for the component and ordering that vertex's neighbor list.  The starting vertex is a vertex in the current component that is connected to the minimum vertex of the enclosing face.  We order the neighbor lists by traveling counterclockwise around the current vertex and we use this ordering to calculate the edges of the slices in a non overlapping manner. \emph{NumberComponent} performs a depth first search on the component, numbering the vertices and ordering their individual neighbor lists.  The numbers for the example graph $G$ given in figure \ref{baker1} are given in figure \ref{baker2}.\par
\begin{figure}[htb] \begin{center}  \input{baker3.pstex_t} \caption{The three main slices for the graph given in figure \ref{baker1}} \label{baker3}\end{center}\end{figure}
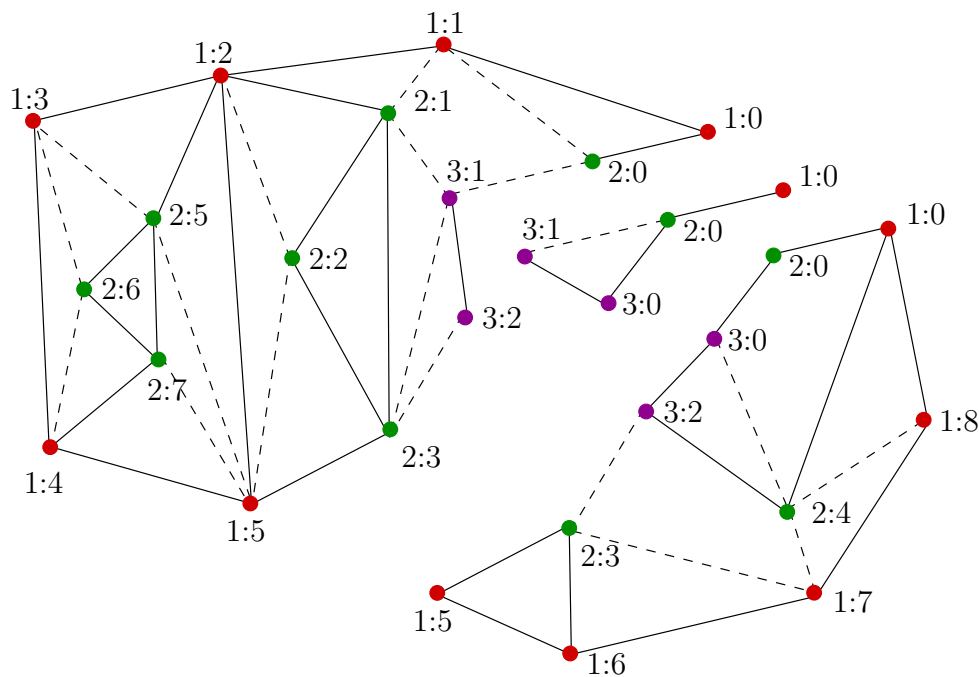

Once the vertices are numbered we break $G'$ into $s$ slices using the methods \emph{MakeSlices} and \emph{ConstructRB}.  The result is a series of $n$ slices (where $n$ is the number of edges in the innermost starting component that we use to construct the slices) that contain all of $G'$, with each slice sharing boundaries with its two neighboring slices.  Figure \ref{baker3} gives the slices for the graph $G'$ shown in figure \ref{baker1}  The component we used to construct the slices contains the set of vertices $\{$3:0, 3:1, 3:2$\}$ thus, we start with three initial slices. \par 
%%%%%%%%%%%%%%Setup
\begin{algorithm}[p]{Setup} - Initialize the vertex numbers to zero\\
\textbf{Global variables:} graph $G'$
\begin{algorithmic}
\FORALL {$v \in V(G')$}
\STATE \num{v} $ := 0$
\ENDFOR
\end{algorithmic}
\end{algorithm}

%%%%%%%%%%%number graph
\begin{algorithm}[p]{NumberGraph} - Number the vertices of the graph working inwards one component at a time\\ 
\textbf{Global variables:} component list $C$ for graph $G'$, integer $i$
\begin{algorithmic}
\STATE $i := 0$
\STATE choose a vertex $v$ in the level 1 component $c$
\STATE NumberComponent $(v, null)$
\STATE remove $c$ from $C$
\FORALL {level $l$ starting with $l = 2$}
\STATE $i := 0$ 
\WHILE{there are components in $C$ with level $l$}
\STATE let $c$ be the level $l$ component containing the level $l$ vertex with the minimum number
\STATE let $u$ be the vertex with the minimum number in the face enclosing $c$
\STATE let $v$ be the first vertex in $\Gamma(u)$ such that $v \in c$
\STATE NumberComponent $(v, u)$
\ENDWHILE

\ENDFOR
\end{algorithmic}
\end{algorithm}
%%%%%%%%%%%number components
\begin{algorithm}[p]{NumberComponent} - Number the vertices in a particular component by traveling counterclockwise around the component\\
\textbf{Input:} vertices $current$, $reference \in V(G')$\\
\textbf{Global variables:} graph $G'$, integer $i$
\begin{algorithmic}
\STATE \num{current} $:= i$
\STATE $i := i + 1$
\IF{for all $u \in \Gamma(current)$, \num{u} $= 0$ $\wedge$ \level{current} $= 1$}
\STATE reorder $\Gamma(current)$ starting with the first counterclockwise neighbor of $current$
\ELSE 
\STATE reorder $\Gamma(current)$ starting with the first counterclockwise neighbor after $reference$
\ENDIF
\FORALL {$u \in \Gamma(current)$}
\IF {\num{u} $= 0$ $\wedge$ \level{u} = \level{v}}
\STATE call NumberComponent $(current, u)$
\ENDIF
\ENDFOR
\end{algorithmic}
\end{algorithm}
%%%%%%%%%%%%%%%%%%%%make slices
\begin{algorithm}[p]{MakeSlices} - Break the graph $G'$ into slices based on the vertex numbers determined by NumberGraph and NumberComponent \\
\textbf{Global variables:} graph $G'$
\begin{algorithmic}
\STATE let $c$ be the innermost component that contains a vertex $v$ with \num{v} $= 1$
\STATE let $i$ = 1
\STATE create a new slice $i$
\STATE \lb{i} = $\{v \mid$  \num{v} $= 0\}$
\STATE ConstructRB($i$)
\STATE let $n = ||\{u \mid u \in c \}||$
\FOR {$i = 2, ..., n-1$}
\STATE create a new slice $i$
\STATE \lb{i} = \Rb{i-1}
\STATE ConstructRB($i$)
\ENDFOR
\STATE $i := n$
\STATE \lb{i} = \Rb{i-1}
\STATE \Rb{i} = \lb{1}
\end{algorithmic}
\end{algorithm}
%%%%%%%%%%%%%%%%%%%%construct rb
\begin{algorithm}[p]{ConstructRB} - Construct the right boundary of the given slice by traversing a path outwards via neighbors with decreasing level numbers\\
\textbf{Input:} slice $s$\\
\textbf{Global variables:} graph $G'$
\begin{algorithmic}
\STATE let $v$ be the innermost vertex in \lb{s}
\STATE let $u$ be the first vertex in $\Gamma(v)$ with \level{u} = \level{v}
\STATE \Rb{s} $:= \{u\}$
\FOR {$l =$ \level{v}, ..., 1}
\STATE let vertex $x$ be the level $l$ vertex in \Rb{s}
\STATE let vertex $y$ be the first vertex in $\Gamma(x)$ with \level{y} $= l-1$
\STATE \Rb{s} $:=$ \Rb{s} $\cup \{y\}$
\ENDFOR
\end{algorithmic}
\end{algorithm}
%%%%%%%%%%%%%%%%%%%%%%%%find center
\begin{algorithm}[p]{FindCenter} - Find the pivot point of a slice for a particular level (we can then break the slice into smaller pieces and merge them together from the left and right of the pivot)\\
\textbf{Input:} slice $s$, level $l$\\
\textbf{Global variables:} graph $G'$
\begin{algorithmic}
\STATE let vertex $u$ be the level $l$ vertex in \Rb{x}
\STATE let vertex $v$ be the level $l+1$ vertex in \lb{s}
\WHILE {edge $uv \notin E(G')$}
\STATE $u :=$ the first counter clockwise neighbor of $u$ in level $l$
\ENDWHILE
\RETURN $u$
\end{algorithmic}
\end{algorithm}

Within each slice $i$, we recursively divide the slice into smaller slices.  For each level, (after the innermost level) we calculate the center point using the method \emph{FindCenter}.  We use the center point of the slice to make thinner, non-overlapping slices that are merged together from the left and the right.  This merging process uses dynamic programming to bound the size of the merged tables; recall that the slice boundaries are bounded by the number of levels in the slice which, during this recursion, is non-increasing.  The smaller slices that make up the largest slice in figure \ref{baker3} are given in figure \ref{baker4}.  The center point for this slice is labeled 2:3.  The smaller slices that make up the the largest slice in figure \ref{baker4} are given in figure \ref{baker5}, the center point in this case is the vertex with label 5:7.  
\begin{figure}[p] \begin{center}  \input{baker4.pstex_t} \caption{The slices that make up the largest slice in figure \ref{baker3}} \label{baker4}\end{center}\end{figure}
\begin{figure}[p] \begin{center}  \input{baker5.pstex_t} \caption{The slices that make up the largest slice in figure \ref{baker4}} \label{baker5}\end{center}\end{figure}
We continue making smaller and smaller slices until we have just a single edge at which point the recursion stops and we create a table for that edge.  We then work our way back towards the center of the graph merging the slices (and their corresponding tables together).  \par
We use the methods \emph{NewTable} and  \emph{MergeTables} for table construction and maintenance.  The tables used here are similar to those described in the previous chapter, but it is useful here to distinguish between the vertices in $s$ that are from the left bound of the slice and those from the right bound.  \emph{NewTable} creates a table given an edge of graph $G'$, with the possible combinations of the endpoints of the edge in the independent set.  \emph{MergeTables} accepts two slices $i$ and $j$ to be merged.  For each set $s \subseteq$ (\lb{i} $\cup$ \Rb{j}), we look for the rows $a$ and $b$ in the tables for $i$ and $j$ respectively, such that \lb{i.a.s} = \lb{s},  \Rb{j.b.s} = \Rb{s}, and \Rb{i.a.s} = \lb{j.b.s}.  The size of the Max IS, including the vertices in $s$, for the new slice is the sum of the independent set sizes for rows $a$ and $b$ minus their overlap (in the shared \Rb{i}, \lb{j} boundary). \par

\begin{algorithm}[p]{NewTable} - Create a new table given a single edge\\
\textbf{Input:} level 1 edge e with endpoints $u$ and $v$\\
\textbf{Global variables: graph $G'$} 
\begin{algorithmic}
\STATE create a new table t
\STATE add row $t.1$ with $t.1.s = \emptyset$, $t.1.is = 0$
\STATE add row $t.2$ with $t.2.s = \{u\}$, $t.2.is = 1$
\STATE add row $t.3$ with $t.3.s = \{v\}$, $t.3.is = 1$
\STATE if $uv \notin E(G)$, add row $t.4$ with $t.4.s = \{u$, $v\}$ and $t.4.is = 2$
\end{algorithmic}
\end{algorithm}

\begin{algorithm}[p]{MergeTables} - Merge tables for slices (with a shared boundary) together (removing entries from the table that contain edges in $G$) \\
\textbf{Input:} slices $i$, $j$ with shared boarder \Rb{i} (\lb{j})\\
\textbf{Global variables} graph $G$
\begin{algorithmic}
\STATE create a new table $t$ with one row for each subset of \lb{i} $\times$ \Rb{j}
\FORALL {row $r$}
\IF {there exists an $x$ and  $y \in t.r.s$ such that $xy \in E(G)$}
\STATE remove row $t.r$
\ELSE 
\STATE set $t.r.s = -1$
\FORALL { rows $i.a$ and $j.b$ where \lb{i.a.s} = \lb{t.r.s} and \Rb{j.b.s} = \Rb{t.r.s}}
\IF { \Rb{i.a.s} = \lb{j.b.s}}
\STATE $t.r.s = \max \{t.r.is$, $i.a.is + j.b.is - ||$\Rb{i.a.s}$|| \}$
\ENDIF
\ENDFOR
\ENDIF
\ENDFOR
\end{algorithmic}
\end{algorithm}

Baker proves the correctness of her algorithm by showing that the recursion is finite and that the tables are computed correctly for each slice.  Since the final slice contains the entire graph, the table for this slice contains the value of the maximum independent set for the graph.\par

Baker's algorithm uses the properties involving separators that we discussed in chapters 3 and 5.
\begin{lemma}\label{bakerlemma1}$(\forall $ slices $i$, $j$, $i \not= j)$  $(\forall~ u$, $v \in V(G))$  $[(u \in$ $(V(i) - ($\lb{i} $\cup$ \Rb{i}$))) \wedge (v \in (V(j) - ($\lb{j} $\cup$ \Rb{j}$))) \Rightarrow uv \notin E(G)]$.\end{lemma}
Proof by contradiction: Assume $(\exists $ slices $i$, $j$, $i \not= j)$  $(\exists~ u$, $v \in V(G))$  such that $(u \in$ $(V(i) - ($\lb{i} $\cup$ \Rb{i}$)))$,  $(v \in (V(j) - ($\lb{j} $\cup$ \Rb{j}$)))$ and $uv \in (G)$. Recall that $f'$ was formed by adding edges to a planar embedding of $G$ such that $G'$ is planar and triangulated.  The vertices in the boundaries of $i$ form a path from the innermost level to level 1.  If $u$ and $v$ are connected this would create an edge crossing.\par
\begin{corollary} The boundaries of the slices are separators of the graph.  \end{corollary}
The boundaries of slice $i$ separate the vertices in $i - ($\lb{i} $\cup$ \Rb{i}$)$ from the vertices in $V(G) - V(i)$.  \par

%%%%%%%%%%%%%%%plane definitions%%%%%%%%%%%%%%%%
\chapter{Plane Decompositions}
In our attempts to extend the tree decomposition definition to allow for cycles (such that the resulting decomposition is still planar), we investigated several plane decomposition definitions.  (We call them plane decompositions rather than planar decompositions because this term has already been used in measuring how close to being planar a particular graph is, see for example J\"unger et al. \cite{junger}.)  We write \pd{X} to denote the set of decompositions of $G$ which conform to the definition labeled PDX.  We denote the planewidth of a graph using a particular decomposition with \pw{X}. Below is a list of criteria by which we evaluated each definition.\par
\begin{enumerate}
\item Non-triviality: for all integers $n$, there exists a graph $G$ with \pw{X} $= n$.
\item Soundness with respect to planarity: Graph $G$ is planar $\Rightarrow$ \pw{X} of $G$ is 1.
\item Completeness with respect to tree decompositions: For graph $G$, decomposition $D \in$ \td{\ref{t1}} $\Rightarrow D \in$ \pd{X}.
\item Soundness with respect to tree decompositions: For graph $G$, decomposition $D \in$ \pd{X} is a tree $\Rightarrow D \in$ \td{\ref{t1}}.
\item Non-trivial difference from tree decompositions: For any $k$, $j \geq k$, there exists a graph $G$ of treewidth $j$ and \pw{X}$=k$.
\item Algorithmic utility: For any $k$, there should be an efficient algorithm for solving NP-Hard problems when restricted to the set of graphs with \pw{X} $= k$
\end{enumerate}

We started our investigation with criteria 3 and 5, which are clear goals for extending Definition TD\ref{t1} to a non-trivial plane decomposition.  We added criteria 1, 2, and 4 to make the plane decomposition definition more robust and complete.  Criterion 6 was our long term goal, although it was not clear until we started studying algorithms for planar graphs how this criterion would contribute to the definition.  Figure \ref{crittab} gives a summary how each definition given below fits with the criteria listed above.   
\begin{figure}[htb]
\begin{tabular}{|c|c|c|c|c|c|c|c|} \hline
criterion & PD\ref{p1} & PD\ref{p2} & PD\ref{p3} & PD\ref{p4} & PD\ref{p5} & PD\ref{p6}& PD\ref{p7}\\ \hline
1 & + & & & + & + & + & + \\ \hline
2 &  & + & + &  &  &  &  \\ \hline
3 & + & + & + &  &  &  &  \\ \hline
4 & + & + & + &  & + & + & + \\ \hline
5 &  & & & + & + & + & + \\ \hline
6 & + & & &  &  &  & + \\ \hline
\end{tabular}\\For each decomposition definition (other than the base definition (PD\ref{p0}) we denote met criteria with a '+' sign.  \caption{Criteria met by possible plane decomposition definitions}\label{crittab}\end{figure}

The first definition gives us a basic decomposition framework from which to build our plane decompositions.  \par
%%%%%%%%%%%%%plane 0 %%%%%%%%%%%%%%%%%%%%%
\begin{plane}\label{p0}Given graph $G$, decomposition $D \in$ \pd{\ref{p0}}  is a pair $(S$, $P)$ with $P$ a connected, planar graph, and  $S = \{X_{i} \mid i \in V(P)\}$ a family of subsets of $V$, one for each node of $P$, such that
\begin{enumerate}
\item $\bigcup_{i \in V(P)} X_{i} = V(G)$,
\item $(\forall~ uv \in E(G))$ $(\exists~ \; i \in V(P))$ $[u$, $v \in X_{i}]$
\end{enumerate}
The width of $D$ is $max_{i \in V(P)} ||X_{i}|| - 1$, and the \pw{\ref{p0}} of a graph $G$ is the minimum width over all plane decompositions of $G$.\end{plane}

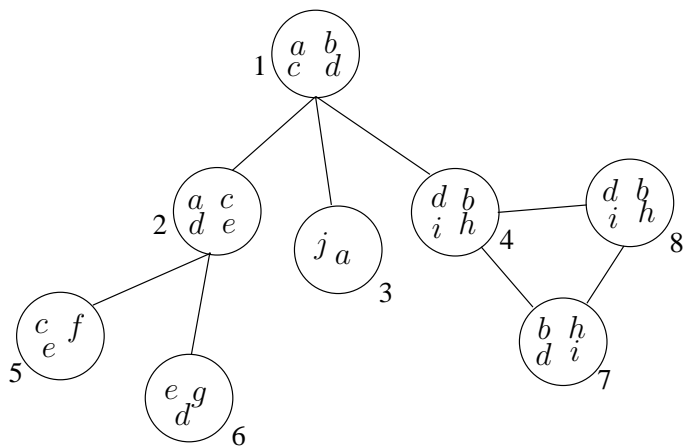
\begin{figure}[htb] \begin{center}  \input{d2.pstex_t} \caption{Decomposition $D \in$ \pd{\ref{p1}} of graph $G$ shown in figure \ref{g1}.} \label{d2}\end{center}\end{figure}

%%%%%%%%%%%%%plane 1
Perhaps the most straightforward modification of definition TD\ref{t1} is to relax the constraints on the decomposition to allow for cycles.
\begin{plane}\label{p1}Given graph $G$ and decomposition $D \in$ \pd{\ref{p0}}, $D \in$ \pd{\ref{p1}} if
\begin{enumerate}
\setcounter{enumi}{2}
\item $(\forall~ i$, $j$, $k \in V(P))$ $[(j$ is on the path from $i$ to $k$ in $P) \Rightarrow (X_{i} \cap X_{k} \subseteq X_{j})]$.
\end{enumerate} \end{plane}
The decomposition shown in figure \ref{d2} is a decomposition $D \in$ \pd{\ref{p1}} of the graph $G$ shown in figure \ref{g1}.  Note that if we removed node 8 we would have a valid tree decomposition of $G$ (we could also remove node 7).  \par
Since the only change we made to TD\ref{t1} was to allow cycles, for decomposition $D$ of graph $G$,  $D \in$ \td{\ref{t1}} $\Rightarrow D \in $ \pd{\ref{p1}}.  Thus, Definition PD\ref{p1} meets criteria 3 and 4; however, it violates criteria 5 as we will show with the following two lemmas.
%%%%%%%%%%%%%%%%% decomp1 = tree decomp
\begin{lemma}\label{p1tree} Given graph $G$ and decomposition $D \in$ \pd{\ref{p1}} with width $k$, there exists a decomposition $D' \in$ \td{\ref{t1}} with width $k$.\end{lemma}
\begin{proof} Let $D = (S, P)$.  We construct $D' = (S', P')$ as follows.  Start with $S' = S$ and $P' = P$.  While $P'$ contains cycles: Let $C$ be a cycle in $P'$, For some $i \in C$, we replace $C$ with node $i$.\\
Now we show that $D' \in$ \td{\ref{t1}}.  Let $R$ be the set of nodes that were removed during the elimination of cycle $C$ and let node $i$ be the node in $C - R$.  Since a cycle is a path starting and ending with the same node, by Definition PD\ref{p1}.3, $(\forall~ j \in R)(\forall~ x \in X_{j})[x \in X_{i}]$.  Thus, since $D \in$ \pd{\ref{p1}} and $P'$ is a tree, $D' \in $ \td{\ref{t1}}.  Also note that for a cycle $C$ in $P$, $(\forall~ i$, $j \in C)[||X_{i}|| = ||X_{j}||]$.  So, the width of $D$ is the same as the width of $D'$. \end{proof}

\begin{lemma}For graph $G$, the \pw{\ref{p1}} of $G$ and the \tw{\ref{t1}} of $G$ are the same.\end{lemma}  
\begin{proof}
Let $D \in$ \pd{\ref{p1}} be a plane decomposition of minimum width and let $D' \in $ \td{\ref{t1}} be a tree decomposition of minimum width. Clearly by a quick examination of Definitions PD\ref{p1} and TD\ref{t1}, we can see that $D' \in $ \pd{\ref{p1}}.  Thus, the width of $D \leq$ the width of $D'$.  From lemma \ref{p1tree} we can construct decomposition $D'' \in$ \td{\ref{t1}} with width equal to that of $D$.  Thus, the widths of $D$ and $D'$ are the same, and by our assumption, the \pw{\ref{p1}} of $G$ and the \tw{\ref{t1}} of $G$ are the same.
\end{proof}

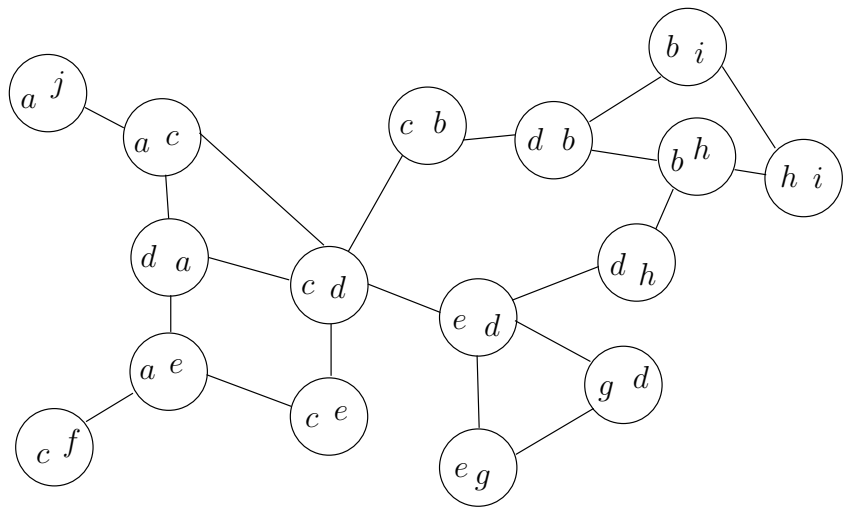
\begin{figure}[htb] \begin{center}  \input{d4.pstex_t} \caption{Decomposition $D \in$ \pd{\ref{p3}} of graph $G$ shown in figure \ref{g1}.} \label{d4}\end{center}\end{figure}

Although Definition PD\ref{p1} would also meet criterion 6 (since it would not be hard to eliminate cycles and use existing polynomial time algorithms for graphs of fixed treewidth), it is not an ideal decomposition.  Our next modification came from relaxing Definition PD\ref{p1}.3 to require that only the nodes on one path (rather than all paths) between two nodes in the decomposition contain their intersection.
%%%%%%%%%%%%%plane 2%%%%%%%%%%%%%%%%%%%%%%%55
\begin{plane}\label{p2} Given graph $G$ and decomposition $D \in $ \pd{\ref{p0}}, $D \in$ \pd{\ref{p2}} if
\begin{enumerate}
\setcounter{enumi}{2}
\item $(\forall~ i$, $k \in V(P))$ $[(X_{i} \cap X_{k} \not= \emptyset) \Rightarrow (\exists$ a path $p$ from $i$ to $k$ in $P)$ $(\forall~ j \in p)$ $(X_{i} \cap X_{k} \subseteq X_{j})]$.
\end{enumerate}\end{plane}
Figure \ref{d4} shows decomposition $D \in $ \pd{\ref{p2}} of graph $G$ shown in figure \ref{g1}.  It is easy to see that PD\ref{p2} also meets criterion 3, and since there is only one path between any two nodes in a tree, it also meets criterion 4.  PD\ref{p2}  fails to meet criteria 1 and 5, which we show with the following lemma.  (Note that although we do not directly prove that all graphs have a \pw{\ref{p2}} of 1, we show that all cliques have a \pw{\ref{p2}} of 1 and cliques are in some sense the ``least planar'' types of graphs).  
%%%%%%%%%%%%% decomp2 trivial
\begin{lemma} Given a clique $G$, there exists a decomposition $D \in$ \pd{\ref{p2}} such that width of $D$ is  $1$.\end{lemma}
\begin{proof} A plane decomposition $D\in$ \pd{\ref{p2}} with width = 1 of a graph $G$, where $G$ is a clique of size $n$, can be constructed of follows. (Assume WLOG that $n$ is odd.)\\
$D = (S, P)$ with $S = \{\{u, v\} \mid uv \in E(G)\}$, and $V(P) = \{0, ..., \left(\!\begin{array}{c}n \\ 2\end{array} \!\right)-1\}$. \\
Let $V(G) = \{v_{0}, ..., v_{n-1}\}$.\\
Construct a cycle of the first $n$ nodes of $I$, numbered 0 through $n-1$, each containing two consecutive vertices of $G$, i.e. $\{v_{0}$, $v_{1}\}$, $\{v_{1}$, $v_{2}\}$, ..., $\{v_{n-2}$, $v_{n-1}\}$, $\{v_{n-1}$, $v_{0}\}$.  
Starting between the first two nodes of the previous cycle, construct another cycle consisting of the next $n$ nodes, numbered $n$ through $2n-1$ each containing two vertices of $G$ following the pattern $\{v_{i}, v_{(i + 2)\bmod n}\}$ for $0 \leq i \leq n-1$, i.e. $\{v_{0}, v_{2}\}, \{v_{1}, v_{3}\}, ..., \{v_{n-2}, v_{0}\}, \{v_{n-1}, v_{1}\}$.
Connect each node $j$ of this second cycle to the two nodes of the first cycle numbered $j\bmod n$ and $j \bmod n + 1$.
The next cycle will have nodes containing vertices numbered $i$, $(i+3)\bmod n$ and will start between the first two nodes of the previous cycle.  Continue in this manner until all edges in $E(G)$ have been added. For example, see figure \ref{circleconst} which shows the decomposition of $K_{5}$.
The path containing vertex $v_{i}$ passes through two nodes from each cycle, numbered $j$ and $j - \lfloor j/n \rfloor -1$, where $i \equiv j \bmod n$.  The nodes containing vertex $v_{3}$ are blue in figure \ref{circleconst}. \end{proof}
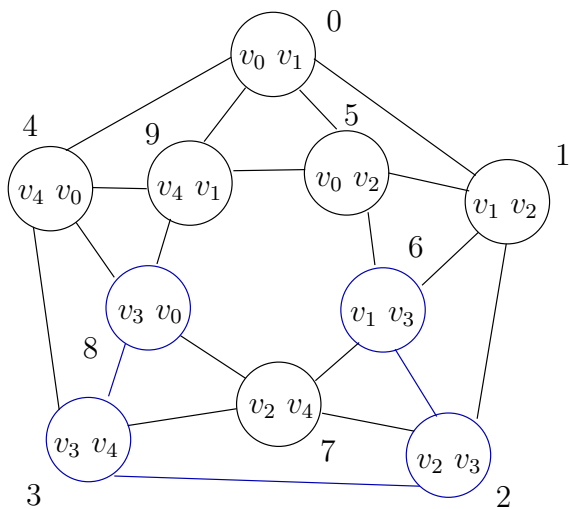
\begin{figure}[htb] \begin{center}  \input{circle_const.pstex_t} \caption{Decomposition $D \in PD\ref{p2}(K_{5})$.} \label{circleconst}\end{center}\end{figure}
The next definition is a strengthening of Definition PD\ref{p2}.3 to require that the nodes on the shortest path between two nodes contain their intersection.
%%%%%%%%%%%%%%%%%%%plane 3%%%%%%%%%%%%%%%%%5555
\begin{plane}\label{p3} Given graph $G$ and decomposition $D \in$ \pd{\ref{p0}}, $D \in$ \pd{\ref{p3}} if
\begin{enumerate}
\setcounter{enumi}{2}
\item $(\forall~ i$, $j$, $k \in V(P))$ $[(j$ is on a shortest path between $i$ and $k$ in $P) \Rightarrow (X_{i} \cap X_{k} \subseteq X_{j})]$.
\end{enumerate} \end{plane}
Figure \ref{d4} is also a decomposition $D \in$ \pd{\ref{p3}} of graph $G$ shown in figure \ref{g1}.  This definition meets criteria 3 and 4, as its predecessors, but it also fails criteria 1 and 5 with a slightly less intuitive construction that is given in the next lemma.

%%%%%%%%%%% decomp 3 trivial
\begin{lemma} Given clique $G$, there exists a decomposition $D \in$ \pd{\ref{p3}} such that width of $D$ is $1$.\end{lemma}
\begin{proof} A decomposition $D \in$ \pd{\ref{p3}}, of graph $G$ with width $=1$ for a clique with $||V(G)|| =n$ can be constructed as follows.\\
Begin with a lower triangular grid of size $(n-1) \times (n-1)$. Label the grid from top to bottom with the numbers $2, ... ,n$ and from left to right with the numbers $1, ..., n-1$.  The lines of the grid represent edges of $D$, and each node of $D$ contains $v_{i}, v_{j} \in V(G)$ where $i$ and $j$ are designated by the labels on the intersecting grid lines.  For example, see figure \ref{triangleconst} which is a decomposition $D \in$ $PD\ref{p3}(K_{5})$.  The blue node represents $X_{i} = \{v_{2}$, $v_{3}\}$.  The green nodes contain the vertex $v_{4}$, a shortest path between any two of these lies along a green edge.\\
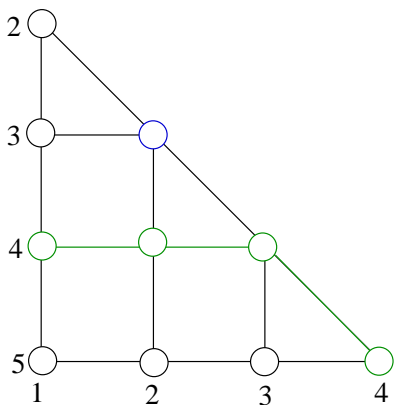
\begin{figure}[htb] \begin{center}  \input{triangle_const.pstex_t} \caption{Decomposition $D \in PD\ref{p3}(K_{5})$ with width of 1.} \label{triangleconst}\end{center}\end{figure}
\noindent We prove that $D \in$ \pd{\ref{p3}} with induction on $n$: The base case, the trivial case of $n = 2$, is true because there are only two vertices $v_{1}$ and $v_{2}$ and one edge connecting them.  The plane decomposition thus consists of one node containing both $v_{1}$ and $v_{2}$.  If $n = 3$, the plane decomposition is a triangle with three nodes containing $\{v_{1}, v_{2}\}, \{v_{2}, v_{3}\}$, and $\{v_{1}, v_{3}\}$.  This is a valid decomposition with a width of 1.\\
Induction hypothesis: For $n = l$, the construction given above results in a valid plane decomposition according to definition PD\ref{p3} of width 1.  Notice that for all $v_{i} \in V$, and nodes $j, k \in I$ where $v_{i} \in X_{j} \cap X_{k}$, a shortest path between $j$ and $k$ exists along the grid lines labeled $i$ and may include the only possible diagonal edge from the row of nodes containing $v_{i}$ to the column of nodes containing $v_{i}$.\\
Let $n=l+1$.  First, we construct a decomposition of the first $l$ vertices as described above, which yields an $l \times l$ triangular grid.  From the induction hypothesis, this upper triangle is a plane decomposition of a click of size $l$.  Consider vertex $v_{l+1} \in V(G)$  We add another row labeled $v_{l+1}$ and a column labeled $v_{l}$; this added column contains only one node which contains $\{v_{l}, v_{l+1}\}$, clearly a shortest path containing any two nodes of this new row is along this row.  Thus, this is a valid plane decomposition with width $=1$.\end{proof}

%%%%%%%%%%%%%%%%%%%%%%%%%%plane 4%%%%%%%%%%%%%%%%%%%%%%
 The next definition is different from the others in that the contents of $E(P)$ (for decomposition $D = (S, P)$) is based on the contents of the bags in $S$.
\begin{plane}\label{p4}Given graph $G$ and decomposition $D \in$ \pd{\ref{p0}}, $D \in$ \pd{\ref{p4}} if
\begin{enumerate}
\setcounter{enumi}{2}
\item $(\forall~ i$, $j \in V(P))$ $[(ij \in F) \Leftrightarrow (\exists~ u \in X_{i}$, $v \in X_{j})$ $[uv \in E(G)]]$.
\end{enumerate}\end{plane}
Figure \ref{d5} shows a decomposition $D \in$ \pd{\ref{p4}} for graph $G$ given in figure \ref{g1}.  Although this definition does not allow for trivial decompositions as described above, it seems unnecessarily restrictive compared to the tree decomposition definition.  It meets criterion 1 but not criteria 3 or 4. The following lemma gives us a lower bound on the \pw{\ref{p4}} of a clique. \par
\begin{figure}[htb] \begin{center}  \input{d5.pstex_t} \caption{Decomposition $D \in$ \pd{\ref{p4}} of graph $G$ shown in figure \ref{g1}.} \label{d5}\end{center}\end{figure}
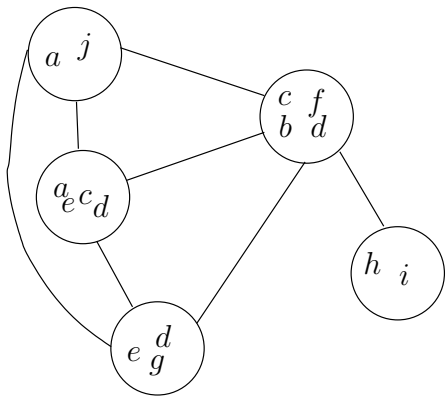
\begin{lemma}\label{cliquewidth d4} For graph $G$, if $G$ is a clique, $G$ has \pw{\ref{p4}} $\geq ||V(G)||/4$. \end{lemma}
\begin{proof} Let $D$ be a decomposition of $G$ of minimum width such that $D \in$ \pd{\ref{p4}}.  Then $(\forall~ i$, $j \in V(P) : i \not= j)(\forall~ u \in X_{i})$ $(\forall~v \in X_{j})[uv \in E(G)]$.  Thus $P$ is a clique and since $K_{4}$ is the largest planar graph that is a clique, the width of $D$ is $\geq ||V(G)||/4$.  Since $D$ is minimum width, the \pw{\ref{p4}} of $G$ is also $\geq ||V(G)||/4$.\end{proof}

The next lemma shows that PD\ref{p4} does not meet criteria 2 and 3.

\begin{lemma}\label{counterex} There exists a graph $G$ with decompositions $D=(S,T)$ and $D'=(S',P')$ such that $D\in$ \td{\ref{t1}}, $D' \in $ \pd{\ref{p4}}, $D \notin$ \pd{\ref{p4}}, and $P'$ is a tree, but $D' \notin$ \td{\ref{t1}}.\end{lemma}
\begin{proof} Let $G$ be the path with $V(G) = \{a$, $b$, $c$, $d\}$ and $E(G) = \{ab, bc, cd\}$.  See figure \ref{counter1}.a. Let $S = \{ \{a, b\}$, $\{b, c\}$, $\{c, d\}\}$, $V(T) = \{1, 2, 3\}$, and $E(T) = \{12, 23\}$.  $D$ is shown in figure \ref{counter1}.b, note that nodes 1 and 3 should be connected for $D$ to be a decomposition according to Definition PD\ref{p4}.  Let $S' = \{\{c, d\}$, $\{a, b, c\}$,  $\{c, d\}\}$, $V(P') = \{1, 2, 3\}$, and $E(P') = \{12, 23\}$. $D'$ is shown in figure \ref{counter1}.c, note that $X_{1}$ and $X_{3}$ contain $d$, but $X_{2}$ does not. Thus $D' \notin$ \td{\ref{t1}}.  \end{proof}
\begin{figure}[htb] \begin{center}  \input{counter1.pstex_t} \\$G$ with decompositions $D$ and $D'$ such that $D\in$ \td{\ref{t1}}, $D' \in $ \pd{\ref{p4}}, $D \notin$ \pd{\ref{p4}}, and $P'$ is a tree, but $D' \notin$ \td{\ref{t1}} \caption{Example graphs for Lemma \ref{counterex}} \label{counter1}\end{center}\end{figure}

Recall from our discussion of tree decompositions and algorithms that use decompositions,  the property relating separators of the decomposition and separators of the graph.  The following two definitions were written with that property in mind.  Definition PD\ref{p5} has a straightforward inclusion of the property discussed in chapter 2 (lemma \ref{sep1}); Definition PD\ref{p6} is an attempt to restate this property in more intuitive terms.  First, we show that the two definitions are the same.  Then, we discuss the criteria for good plane decompositions in relation to Definition PD\ref{p6}.
%%%%%%%%%%%%%plane 5%%%%%%%%%%%%%%%%%%%
\begin{plane}\label{p5}Given graph $G$ and decomposition $D \in$ \pd{\ref{p0}}, $D \in$ \pd{\ref{p5}} if
\begin{enumerate}
\setcounter{enumi}{2}
\item $\forall~$ separators $J \subseteq V(P)$, let $C_{1}, ..., C_{n}$ be the connected components of $P \; \backslash \;J$,  $(\forall~ C_{i})[(\bigcup_{j \in C_{i}}X_{j} - \bigcup_{j \in J}X_{j}) \not= \emptyset]$  $\wedge$ $(\forall~ a$, $b \in \{1, ..., n\}: a \not= b)$ $([(\bigcup _{i \in C_{a}}X_{i} - \bigcup_{k \in J}X_{k})$ $\cap$ $(\bigcup _{i \in C_{b}}X_{i} - \bigcup_{k \in J}X_{k}) = \emptyset]$ $\wedge$ $((\forall~ x \in \bigcup_{k \in C_{a}} X_{k})$ $(\forall~ y \in \bigcup_{k \in C_{b}}X_{k})$ $(\{x,y\} \cap \bigcup_{j \in J} X_{j} = \emptyset) \Rightarrow (xy \notin E(G))])$.
\end{enumerate}\end{plane}
%%%%%%%%%%%%%%%%%%%plane 6%%%%%%%%%%%%%%%%%%%%
\begin{plane}\label{p6}Given graph $G$ and decomposition $D \in$ \pd{\ref{p0}}, $D \in$ \pd{\ref{p6}} if
\begin{enumerate}
\setcounter{enumi}{2}
\item $(\forall~ J \subseteq V(P))$ $[(\emptyset \not\subseteq\Gamma(\bigcup_{j \in J}X_{j} - \bigcup_{i \in \Gamma(J)}X_{i}) \subseteq \bigcup_{i \in \Gamma(J) \cup J}X_{i})$  $\wedge$ $((\bigcup_{j \in J}X_{j} - \bigcup_{i \in \Gamma(J)}X_{i}) \cap (\bigcup_{i \in \bar J}X_{i}-\bigcup_{i \in \Gamma(J)}X_{i}) = \emptyset)]$
\end{enumerate}\end{plane}

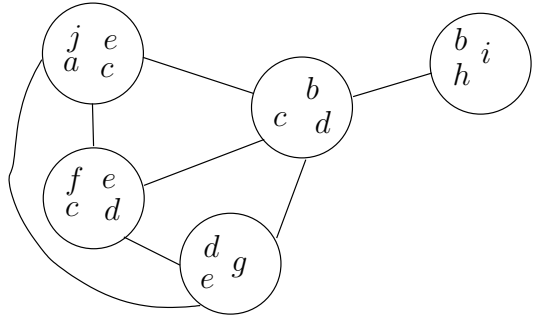
\begin{figure}[htb] \begin{center}  \input{d6.pstex_t} \caption{Decomposition $D \in$ \pd{\ref{p6}} of graph $G$ shown in figure \ref{g1}.} \label{d6}\end{center}\end{figure}

Figure \ref{d6} gives a decomposition $D \in$ \pd{\ref{p6}} for graph $G$ shown in figure \ref{g1}.  

%%%%%%%%%%%%%%%%%%% 5 is a subset of 6
\begin{lemma}For graph $G$, decomposition $D \in$ \pd{\ref{p5}} $\Rightarrow$ $D \in$ \pd{\ref{p6}}. \end{lemma}
\begin{proof}
Let $G$ be a graph, and let $D = (S, P)$ be a plane decomposition of $G$ such that $D \in$ \pd{\ref{p5}}. There are several cases we must consider.\\
First, assume that there exists a $J \subseteq V(P)$ such that $\Gamma(\bigcup_{j \in J} X_{j} - \bigcup_{i \in \Gamma(J)} X_{i}) = \emptyset$.  Let $J_{new} = \Gamma(J)$, then $J$ is a connected component of $P \; \backslash \; J$ and by definition PD\ref{p5}.3 $(\bigcup _{j \in J} X_{j} - \bigcup_{i \in J_{new}} X_{i}) \not= \emptyset$ which causes a contradiction.\\
Next, assume that there exists a $J \subseteq V(P)$ and that there exists an $x$ and $y \in V(G)$ such that $x \in (\bigcup_{j \in J}X_{j} - \bigcup_{i \in \Gamma(J)}X_{i})$, $y \in \Gamma(x)$, but $ y \notin \bigcup_{i \in \Gamma(J) \cup J}X_{i}$.  In other words, $x$ is in $J$ and $y$ is a neighbor of $x$, but $y$ is not in $J$ nor is $y$ in the neighborhood $J$.  See figure \ref{56proof1}.  Let $J_{new} = \Gamma(J)$.  Since $x \in (\bigcup_{j \in J}X_{j} - \bigcup_{i \in \Gamma(J)}X_{i})$, for all $i \in J_{new}$, $x \notin X_{i}$. Also note that for all $i \in  J_{new}, y \notin X_{i}$. From Definition PD\ref{p5}.1, $y$ must be in some $X_{k}$, for $k \in V(P)$.  Consider the connected components of $P \; \backslash \; J_{new}$, let $C_{a}$ = $J$ and let $C_{b}$ be the component that contains $k$.  Then from Definition PD\ref{p5}.3, because $x$ and $y$ are in separate components of $P \; \backslash \; J_{new}$, and there does not exist $i \in J_{new}$ such that $x \in X_{i}$ or $y \in X_{i}$, then $xy \notin E(G)$, which is a contradiction. 
\begin{figure}[htb] \begin{center}  \input{56proof1.pstex_t} \caption{Decomposition $D \in$ \pd{\ref{p5}} of graph $G$.} \label{56proof1}\end{center}\end{figure}\\
Finally assume that there exists a $J \subseteq V(P)$ and that there exists an $a$ and $b$ such that $a \in (J - \Gamma(J))$ and $b \in (\bar J - \Gamma(G))$ and that there exists an $x \in V(G)$ such that $x \in X_{a}$ and $x \in X_{b}$ and $x \notin \bigcup_{j \in \Gamma(J)}X_{j}$.  Let $J_{new}$ be $\Gamma(J)$, from definition PD\ref{p5}.3, because $a$ and $b$ are in separate connected components, we know that $(X_{a} - \bigcup_{k \in J_{new}}X_{k})$ $\cap$ $(X_{b} - \bigcup_{k \in J_{new}}X_{k}) = \emptyset$ which is a contradiction.
\end{proof}
%%%%%%%%%%%%%%%%%%% 6 in 5
\begin{lemma}For graph $G$, decomposition $D \in$ \pd{\ref{p6}} $\Rightarrow$ $D \in$ \pd{\ref{p5}}. \end{lemma}
\begin{proof}
Let $G$ be a graph, and let $D = (S, P)$ be a plane decomposition of $G$ such that $D \in$ \pd{\ref{p6}}.  Again we have several cases to consider.\\
Assume that there exists a separator $J \subseteq V(P)$ with connected components $C_{1}, ..., C_{n}$ of $P \;\backslash\; J$, and that there exists a $C_{i}$ such that $(\bigcup_{j \in C_{i}}X_{j} - \bigcup_{j \in J}X_{j}) = \emptyset$.  Let $J_{new} = \{C_{i}\}$, then by Definition PD\ref{p6}.3,  $\emptyset \not\subseteq \Gamma(\bigcup_{j \in J_{new}}X_{j} - \bigcup_{i \in \Gamma(J_{new})}X_{i})$ which causes a contradiction.\\
Next, assume that there exists a $J \subseteq V(P)$ with connected components $C_{1}, ..., C_{n}$ of $P \;\backslash\; J$, and that there exists a $C_{a}$ and $C_{b}$ (with $a \not= b)$ such that $(\bigcup _{i \in C_{a}}X_{i} - \bigcup_{k \in J}X_{k})$ $\cap$ $(\bigcup _{i \in C_{b}}X_{i} - \bigcup_{k \in J}X_{k}) \not= \emptyset$.  Let $J_{new} = \{C_{a}\}$.  Then by Definition PD\ref{p6}.3, $(\bigcup_{j \in J_{new}}X_{j} - \bigcup_{i \in \Gamma(J_{new})}X_{i}) \cap (\bigcup_{i \in \bar J_{new}}X_{i}-\bigcup_{i \in \Gamma(J_{new})}X_{i}) = \emptyset$ which causes a contradiction.\\
Finally, assume that there exists a $J \subseteq V(P)$ with connected components $C_{1}, ..., C_{n}$ of $P \;\backslash\; J$, and that $(\exists~a$, $b : a \not= b)$ $(\exists~x \in \bigcup_{k \in C_{a}} X_{k})$ $(\exists~y \in \bigcup_{k \in C_{b}} X_{k})$ such that $(\{x$, $y\} \cap \bigcup_{j \in J}X_{j} = \emptyset$) $\wedge$  $(xy \in E(G))$.  See figure \ref{56proof2}.  
Let $J_{new} = C_{a}$, then $ x \in X_{i}$ for some $i \in J_{new}$ and $x \notin \bigcup_{j \in \Gamma(J_{new})}X_{j}$ because the only neighbor of $J_{new}$ is $J$ and by our assumption $x,y \notin \bigcup_{k \in J} X_{k}$.  This is a contradiction because $y \in \Gamma(x)$, but $y \notin \bigcup_{i \in \Gamma(J_{new})}X_{i}$.\end{proof}
\begin{figure}[htb] \begin{center}  \input{56proof2.pstex_t} \caption{Decomposition $D \in$ \pd{\ref{p6}} of graph $G$.} \label{56proof2}\end{center}\end{figure}

We show that Definition PD\ref{p6} meets criteria 1 and 5 below.  Because we used the property involving separators that we introduced in chapter 3, it is clear that Definition PD\ref{p6} meets criteria 4.  This definition does not meet criterion 2, which we also show below.  
\begin{lemma} If graph $G$ is a clique, the \pw{\ref{p6}} of $G$ is at least $||V(G)||/4$.  \end{lemma}
\begin{proof} Let $D$ be a decomposition of $G$ with minimum width such that $D \in$ \pd{\ref{p6}} with $D = (S, P)$.  Assume $P$ is not completely connected, so there exists an $i$, $j$ such that $ij \notin E(P)$.  Let $J = \{i\}$, then from Definition PD\ref{p6}, there exists an $x \in X_{i}$ such that $x \notin \bigcup_{i \in \Gamma(J)}X_{i}$.  Also from Definition PD\ref{p6}, there exists a $y \in X_{j}$ such that $y \notin \bigcup_{i \in \Gamma(J)}X_{i}$ (otherwise if we were to let $J = \{j\}$, $(\bigcup_{j \in J}X_{j} - \bigcup_{i \in \Gamma(J)}X_{i})$ would be empty).  This causes a contradiction because $y \in \Gamma(x)$, but $y \notin \bigcup_{i \in \Gamma(J) \cup J}X_{i}$.  Therefore $P$ must be a clique and from Definition PD\ref{p6}.3, the width of $G$ is $\geq |V(G)|/4$.  Since $D$ has minimum width, \pw{\ref{p6}} of $G$ is at least $||V(G)||/4$. \end{proof}

\begin{lemma} If graph $G$ contains a path of length $\geq 4$, the \pw{\ref{p6}} of $G$ is $> 1$.  \end{lemma}
\begin{proof}
Let $D = (S$, $P)$ be a decomposition of $G$ such that $D \in $ \pd{\ref{p6}} and the width of $D$ is 1.  Let $p$ be a path in $G$ with length $\geq 4$.  Let $v$ be a vertex in $p$ such that $v$ is not an endpoint of $p$.  Let $u$ be a neighbor of $v$ also in $p$ and also not an endpoint of $p$.  Since $u$ and $v$ are neighbors, there exists a $k \in V(P)$ with $u, v \in X_{k}$ from Definition PD\ref{p6}.2.  Let $J = \{k\}$.  Note that because $u$ and $v$ are not endpoints of $p$, $||\{i \in V(P) \mid u \in X_{i}\}|| > 1$ and $||\{i \in V(P) \mid v \in X_{i}\}|| > 1$.  If $\{i \in V(P) \mid u \in X_{i}\}$ $\cap$ $\Gamma(J)  \not= \emptyset$ and $\{i \in V(P) \mid v \in X_{i}\}$ $\cap$ $\Gamma(J) \not= \emptyset$, there is a violation of Definition PD\ref{p6}.3 because $\emptyset \not\subseteq\Gamma(\bigcup_{j \in J}X_{j} - \bigcup_{i \in \Gamma(J)}X_{i})$.  If $\{i \in V(P) \mid u \in X_{i}\} \cap \Gamma(J) = \emptyset$ or $\{i \in V(P) \mid v \in X_{i}\} \cap \Gamma(J) = \emptyset$, this also causes a violation because $(\bigcup_{j \in J}X_{j} - \bigcup_{i \in \Gamma(J)}X_{i}) \cap (\bigcup_{i \in \bar J}X_{i}-\bigcup_{i \in \Gamma(J)}X_{i}) = \emptyset$.
\end{proof}

Recall that Baker's algorithm approximates the Max IS of a graph $G$ by creating a series of graphs $G_{0}, ..., G_{k-1}$ where $G_{i}$ has the levels $\{l \mid l \equiv i \bmod(k+1)\}$ removed for some chosen positive integer $k$.  By the pigeonhole principal we know that one of these graphs has a maximum of $1/(k+1)$ vertices from the maximum independent set.  Definition PD\ref{p6} has the important property that separators of the decomposition are also separators of the underlying graph, but it does not put any restrictions on how many levels of the decomposition can contain a given vertex of the underlying graph.  Thus, removing nodes (and the vertices in their corresponding bags) at levels $\{l \mid l \equiv i \bmod(k+1)\}$ creates a set of graphs with disjoint pieces of the Max IS, but it is possible that elements of the Max IS may appear in every level and thus may never be considered as part of the Max IS.  The next definition solves this problem by placing restrictions on the number of levels in the decomposition a particular vertex may appear in.

Before we give the next plane definition we give a bit of notation.  For graph $G$ and decomposition $D = (S, P)$,  for $v \in V(G)$, let $P_{v}$ be the subgraph of $P$ induced by $\{i \in V(P) \mid v \in X_{i}\}$.  The following definition is based on our previous definition, PD\ref{p6}, but gives a more strict requirement for the location of the  bags containing a vertex $v$ so that we can use Baker's algorithm \cite{baker} to meet criterion 6. 
\begin{plane}\label{p7}Given graph $G$ and decomposition $D \in$ \pd{\ref{p6}}, $D \in$ \pd{\ref{p7}} if
\begin{enumerate}
\setcounter{enumi}{3}
\item There exists a planar embedding $f$ of $P$ such that $(\forall~v \in V(P))$ $[P_{v}$ lies on a single face in $f]$
\end{enumerate}\end{plane}
Because Definition PD\ref{p7} is a more restrictive version of Definition PD\ref{p6}, it also meets criteria 1 and 4 but fails criteria 2 and 3.  The example decomposition given in figure \ref{d6} is also a valid decomposition according to Definition PD\ref{p7}.  Note that if we were to add vertex $c$ to either of the bags it does not currently appear in, we would still have a valid decomposition according to Definition PD\ref{p6} but not Definition PD\ref{p7}.  

\chapter{Approximations for Max IS Using Plane Decompositions}

As we noted above, Baker's algorithm approximates the Max IS of a graph $G$ by finding the Max IS of the subgraphs $G_{0}, ..., G_{k-1}$ where $G_{i}$ has the levels $\{l \mid l \equiv i \bmod(k+1)\}$ removed for some chosen positive integer $k$.  By removing the vertices at various levels we create a limited number of subgraphs each with a bounded number of levels.  Because these subgraphs are disconnected from each other, we can find maximum independent sets of each of them, the union of which will still be an independent set in the original graph.  \par
We can modify Baker's algorithm to work with plane decomposition PD\ref{p7}.  Let $D$ be a decomposition of graph $G$ with $D \in$ \pd{7} and let $f$ be an embedding satisfying Definition PD\ref{p7}.4.  We use the term level 1 vertex (for $v \in V(G)$) to refer to a vertex that appears only in level 1 nodes in $f$.  A level $l$ vertex (for $l > 1$) is a vertex that appears in a level $l$ node in $f$ but not any level $l-1$ node.  Because by Definition PD\ref{p7}.4 all the nodes containing a vertex $v$ must be on a single face, removing the level $l$ nodes of $D$ completely separates the underlying graph $G$ as we show with the following lemma.  
\begin{lemma} \label{sep} Given graph $G$, decomposition $D \in$ \pd{\ref{p7}}, and embedding $f$ of $D$ with $k$ levels, for all levels $l \in \{2, ..., k-1\}$, the level $l$ vertices are a separator of $G$ such that level $i$ vertices for $i > l$ and the level $j$ vertices for $j < l$ do not appear in the same components.\end{lemma}
\begin{proof} Let $G$ be a graph and let $D = (S$, $P)$ be a decomposition such that $D \in$ \pd{\ref{p7}} and let $f$ be an embedding of $D$ with $k$ levels.  Let $l$ be an integer in $\{2, ..., k-1\}$.  Assume that there exists $u$ and $v \in V(G)$ such that $u$ is a level $l-1$ vertex and $v$ is a level $l+1$ vertex and $uv \in E(G)$.  See figure \ref{levels}.  Note that from Definition PD\ref{p7}.3 for every level $i$ (for $1 \leq i \leq k$) there is at least one level $i$ vertex in $G$.  Also note from Definition PD\ref{p7} that $u$ cannot be in a level $l+1$ node and $v$ cannot be in a level $l-1$ node.  From definition PD\ref{p7}.2 if $uv \in E(G)$, then there exists an $i \in V(P)$ such that $\{u, v\} \subseteq X_{i}$.  This causes a contradiction because $v$ is a level $l+1$ vertex.  Referring back to the figure, $u$ can appear in the green nodes, and $v$ in the blue nodes, in order for there to be an edge between them they would need to appear in the same node.  
\end{proof}
\begin{figure}[htb] \begin{center}  \input{levels.pstex_t}\\If vertex $u$ is a level $l-1$ vertex and vertex $v$ is a level $l+1$ vertex, $uv$ cannot be an edge in $E(G)$. \caption{Drawing for Lemma \ref{sep}} \label{levels}\end{center}\end{figure}

Thus, in order to modify Baker's algorithm to use plane decompositions we need only modify the methods \emph{NewTable} and \emph{MergeTables} to consider combinations of vertices of the underlying graph contained in the nodes along the edges of the slices in the decomposition.  The new methods are \emph{PlaneNewTable} and \emph{PlaneMergeTables}.\par
\begin{algorithm}[p]{PlaneNewTable} - Create a new table given a single edge\\
\textbf{Input:} level 1 edge e with endpoints $u$ and $v$\\
\textbf{Global variables: graph $G'$} 
\begin{algorithmic}
\STATE create a new table t
\FORALL {set $s$ in $X_{u} \times X_{v}$}
\IF {$s$ does not contain an edge in $E(G)$}
\STATE add a row $i$ to $t$ with $t.i.s = s$ and $t.i.is = ||s||$
\ENDIF
\ENDFOR
\end{algorithmic}
\end{algorithm}

\begin{algorithm}[p]{PlaneMergeTables} - Merge tables for slices (with a shared boundary) together (removing entries from the table that contain edges in $G$) \\
\textbf{Input:} slices $i$, $j$ with shared boarder \Rb{i} (\lb{j})\\
\textbf{Global variables} graph $G$
\begin{algorithmic}
\STATE create a new table $t$ with one row for each subset of $\bigcup_{m \in lb(i)} X_{m} \times$ $\bigcup_{n \in rb(j)}X_{m}$
\FORALL {row $r$}
\IF {there exists an $x$ and  $y \in t.r.s$ such that $xy \in E(G)$}
\STATE remove row $t.r$
\ELSE 
\STATE set $t.r.s = -1$
\FORALL { rows $i.a$ and $j.b$ where \lb{i.a.s} = \lb{t.r.s} and \Rb{j.b.s} = \Rb{t.r.s}}
\IF { \Rb{i.a.s} = \lb{j.b.s}}
\STATE $t.r.s = \max \{t.r.is$, $i.a.is + j.b.is - ||$\Rb{i.a.s}$|| \}$
\ENDIF
\ENDFOR
\ENDIF
\ENDFOR
\end{algorithmic}
\end{algorithm}
Since our modification of Baker's algorithm changes only how the tables are calculated and not how the slices are constructed the following two theorems follow directly from theorems \ref{bakerth1} and \ref{bakerth2}. 

\begin{theorem} Let $k$ be a positive integer.  Given a $k$-outerplanar embedding of a $k$-outerplanar graph, an optimal solution for maximum independent set can be obtained in time $O(8^{km}n)$, where n is the number of nodes in the decomposition and $m$ is the \pw{\ref{p7}} of $G$.\end{theorem}
\begin{proof}
Baker proves that the number of calls required to build the slices (and thus the tables) is bounded by the number of edges in $G'$, and since $G'$ is planar, the number of edges is linear in the number of vertices.  The merge operation, which is the bulk of the computation, requires $O(8^{km}n)$ time where $n = ||V(P)||$ and $m$ is the \pw{\ref{p7}} of $G$.  (There are a maximum $2^{2km}$ rows in the new table, and for each of these we may have to check up to $2^{km}$ combinations of vertices on the shared boarder between the slices we are merging.)  
\end{proof}
\begin{theorem} For fixed $k$, there is an $O(8^{km}kn)$-time algorithm for maximum independent set that achieves a solution of size at least $k/(k+1)$ optimal for general planar graphs.  Choosing $k = \lceil c~log~log~n \rceil$, where $c$ is a constant, yields an approximation algorithm that runs in time $O(n(\log n)^{3cm}log~log~n)$ and achieves a solution of size at least $\lceil c~log~log~n\rceil / (1 + \lceil c~log~log~n\rceil)$ optimal.  In each case, n is the number of nodes in the decomposition and and $m$ is the \pw{\ref{p7}} of $G$.\end{theorem}
Recall from our discussion of Baker's algorithm that we can use the pigeonhole principle to show that one of the graphs $G_{0}, ..., G_{k-1}$ where $G_{i}$ has the levels $\{l \mid l \equiv i \bmod(k+1)\}$ removed for some chosen positive integer $k$ has a maximum of $1/(k+1)$ vertices from the maximum independent set.  From Lemma \ref{sep} we can also use the pigeonhole principle to show that we will find a $k/(k+1)$ optimal value for the size of the Max IS.
Because we have changed only how the tables are calculated, we assume that recursion is finite and that the slices are formed correctly.  Thus, we need only prove that the tables created by the \emph{PlaneNewTable} and \emph{PlaneMergeTable} methods give correct values for the size of the maximum independent sets of their corresponding slices.  Since the final slice includes the entire graph, we show by induction that the tables are correct starting with the table corresponding to a single edge.  The following lemma gives us this result.
\begin{lemma} The table corresponding to a given slice contains the size of the maximum independent set for the vertices contained in that slice.
\end{lemma}
\begin{proof}
We assume that the \emph{PlaneNewTable} and \emph{PlaneMergeTable} methods are implemented correctly.  If the slice contains just a single edge, then the table is computed by \emph{PlaneNewTable} and contains a row for every combinations of the vertices contained in the endpoints of the edge.  When two slices $i$ and $j$ are merged together we consider every combination of vertices on the boundaries for inclusion in the independent set.  Because the slice boundaries are separators of the decomposition, from Definition PD\ref{p6}.3, vertices that appear only in the interior of slice $i$ cannot be connected to vertices that appear only in the interior of slice $j$.  Thus, when the tables are merged, we can add the values in rows $i.a$ and $b.j$ without creating an invalid independent set.
\end{proof}
We implemented both Baker's algorithm and our modified Baker's algorithm, a discussion of that implementation can be found in Appendix A.
\chapter{Conclusion}
The benefits of tree decompositions are well known.  Given a tree decomposition of a graph we are able to compute solutions to many NP-hard problems efficiently (in linear time).  These dynamic programming algorithms build tables from the decomposition graph, a tree, starting with the leaf nodes.  Because nodes of the tree decomposition are both separators of the decomposition and of the underlying graph (see Lemma \ref{treeclique}), we are able to bound the size of these tables with the width of the decomposition.  \par
Our goal in developing a plane decomposition definition was to combine a more intuitive decomposition from an interpretation viewpoint (by allowing cycles) with a decomposition that was also powerful algorithmically.  After investigating several straightforward modifications of Definition TD\ref{t1} we discovered that the relationship between separators of the decomposition and separators of the underlying graph is key in developing an algorithm for use with our plane decomposition.  This property allows us to use Baker's method \cite{baker} of dividing the graph into slices (with borders of bounded size) and then to compute tables for each of these slices.  Because the slice borders are bounded, the table size is also bounded.  Although this property relating the separators of the graph and its decomposition is strong enough for us to use Baker's algorithm to find independent sets of the graph efficiently, it is not enough for us to make a strong claim about the relationship of the maximum independent set we find compared to the maximum independent set of the graph.  \par
This leads us to our final plane decomposition definition, Definition PD\ref{p7}.  By requiring that vertices appear only in bags on a single face of an embedding of the decomposition, we are able to strengthen our claim about the size of the maximum independent set that we find with our algorithm.  We modify Baker's algorithm to consider all embeddings of the vertices in the bags of the decomposition.  Thus, our tables are bounded in size by the number of levels in the graph and by the size of the bags, and we are able to find a Max IS that is $k/(k+1)$ optimal for some fixed $k$. Definition PD\ref{p7} also meets criteria 1 and 4 but fails criteria 2 and 3.  (The criteria were given in chapter 6).  \par
There are clearly many opportunities for further research.  Algorithms exists which find tree decompositions of width $\leq k$ of graphs for a fixed $k$.  Ideally, we would like to be able to find similar algorithms that find plane decompositions of width $\leq k$ of graphs for a fixed $k$.  Also, it is not clear to us that definition PD\ref{p7} is the most ideal definition.  Although it meets several of the criteria we gave at the start of chapter 6, it does not meet all of them.  Also we are interested to know if Definition PD\ref{p7}.3 can be relaxed without having a negative effect on its algorithmic utility.

%\addcontentsline{toc}{section}{References} 
\bibliographystyle{plain}
\bibliography{plane}

\appendix
\chapter{Implementation}
We implemented Baker's original algorithm \cite{baker} and then modified it as described in chapter 7 to use plane decompositions.  We used Java 1.5 as the implementation language because our goal was to develop a visualization for these algorithms.  As such, we implemented the algorithm for a single $k$-outerplanar graph (as we discussed in chapters 5 and 7).  We followed Baker's algorithm for the construction of the slices, but did not use her methods for constructing the tables.  Baker uses separate methods to create and clean the tables of sets that are not independent.  For graph $G$ and entry $t.i.s$ in table $t$, if $uv \in E(G)$ then $u$ and $v$ can not both be in $t.i.s$.  When we were building table $t$ we checked each entry $t.i.s$ for edges before computing entry $t.i.is$.  Because the bulk of the time spent in this algorithm is spent on the computation of the $is$ entries, it is doubtful that this change had a significant effect on the running time.  \par
The code for this project can be downloaded from\\ http://www.cs.rit.edu/$\sim$mja3749/project.html.  The program can be run by executing the command ``java -jar planeApprox.jar.''  When the program executes two windows appear, one of these is the window containing the decomposition graph.  This decomposition embedding may contain edge crossings, this is because we chose a simple method for placing the nodes in the embedding, by rotating the components and/or spreading the faces and components further apart this embedding would be planar.  The other window displays the table which corresponds to the current slice.  There are three columns in this table.  The first  column shows which bags from the slice borders we are taking vertices from, the second column shows the set of vertices that are being included in the independent set, and the third column shows the size of the Max IS containing these vertices.  \par
We give an example graph $G$ in figure \ref{algg} with a plane decomposition (using definition PD\ref{p7}) in figure \ref{algd}. \begin{figure}[htb] \begin{center}  \input{alg_g.pstex_t}\caption{Graph $G$} \label{algg}\end{center}\end{figure}
\begin{figure}[htb] \begin{center}  \input{alg_d.pstex_t}\caption{Decomposition $D \in$ \pd{\ref{p7}} of Graph $G$ given in figure \ref{algg}} \label{algd}\end{center}\end{figure}
 Information (such as adjacency matrices) about these graphs is entered into the ``test.java'' file provided in the jar file.  An example of the program execution is given in figures \ref{table1} and \ref{drawing1}.  The drawing shows edges that have been covered in green and the edges of the current slice in red.  The slice currently being considered has the borders 1:3 and 1:1 an also contains the node 1:2.  The table given shows the possible independent sets of the vertices contained in the border nodes.  The third row of the table contains the set $\{4\}$ from the original graph (which is contained in the bag labeled 1:3).  The largest independent set from the set $\{4, 13\}$ which contains the vertex 4 has size 1.  Looking at the fourth row, the size of the maximum independent set containing the vertex 3 is 2 (meaning that $\{13, 3\}$ is the largest independent set from this slice containing the vertex $3$).  At the top of the drawing window is a button labeled `next.'  Clicking this button will advance the algorithm to the next slice.  When the algorithm has finished and the final slice contains the entire graph, the size of the maximum independent set of vertices from the underlying graph will appear in the table.  
\begin{figure}[htb] \begin{center}  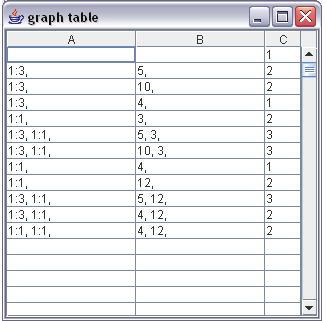\\The above table corresponds to the slice with borders 1:3 and 1:1 from the decomposition shown in figure \ref{algd}. \caption{Table from the execution of planeApprox} \label{table1}\end{center}\end{figure}
\begin{figure}[htb] \begin{center}  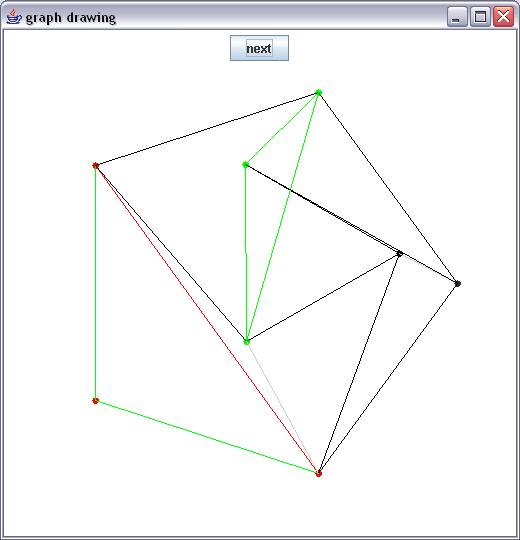\\ The above drawing corresponds to the slice with borders 1:3 and 1:1 from the decomposition shown in figure \ref{algd}. \caption{Drawing from the execution of planeApprox} \label{drawing1}\end{center}\end{figure}
\end{document}

%% file: k4.pstex_t
\begin{picture}(0,0)%
\includegraphics{k4.pstex}%
\end{picture}%
\setlength{\unitlength}{3947sp}%
\begingroup\makeatletter\ifx\SetFigFont\undefined%
\gdef\SetFigFont#1#2#3#4#5{%
  \reset@font\fontsize{#1}{#2pt}%
  \fontfamily{#3}\fontseries{#4}\fontshape{#5}%
  \selectfont}%
\fi\endgroup%
\begin{picture}(6032,1816)(69,-1097)
\put(1265,549){\makebox(0,0)[lb]{\smash{{\SetFigFont{12}{14.4}{\rmdefault}{\mddefault}{\updefault}$b$}}}}
\put(2255,566){\makebox(0,0)[lb]{\smash{{\SetFigFont{12}{14.4}{\rmdefault}{\mddefault}{\updefault}$a$}}}}
\put(5508, 19){\makebox(0,0)[lb]{\smash{{\SetFigFont{12}{14.4}{\rmdefault}{\mddefault}{\updefault}$d$}}}}
\put(5918,540){\makebox(0,0)[lb]{\smash{{\SetFigFont{12}{14.4}{\rmdefault}{\mddefault}{\updefault}$b$}}}}
\put(5508,-587){\makebox(0,0)[lb]{\smash{{\SetFigFont{12}{14.4}{\rmdefault}{\mddefault}{\updefault}$c$}}}}
\put( 69,608){\makebox(0,0)[lb]{\smash{{\SetFigFont{12}{14.4}{\rmdefault}{\mddefault}{\updefault}$a$}}}}
\put( 78,-595){\makebox(0,0)[lb]{\smash{{\SetFigFont{12}{14.4}{\rmdefault}{\mddefault}{\updefault}$c$}}}}
\put(2272,-553){\makebox(0,0)[lb]{\smash{{\SetFigFont{12}{14.4}{\rmdefault}{\mddefault}{\updefault}$c$}}}}
\put(3553,-493){\makebox(0,0)[lb]{\smash{{\SetFigFont{12}{14.4}{\rmdefault}{\mddefault}{\updefault}$d$}}}}
\put(3561,515){\makebox(0,0)[lb]{\smash{{\SetFigFont{12}{14.4}{\rmdefault}{\mddefault}{\updefault}$b$}}}}
\put(4671,515){\makebox(0,0)[lb]{\smash{{\SetFigFont{12}{14.4}{\rmdefault}{\mddefault}{\updefault}$a$}}}}
\put(1222,-544){\makebox(0,0)[lb]{\smash{{\SetFigFont{12}{14.4}{\rmdefault}{\mddefault}{\updefault}$d$}}}}
\end{picture}%

%% file: c6.pstex_t
\begin{picture}(0,0)%
\includegraphics{c6.pstex}%
\end{picture}%
\setlength{\unitlength}{3947sp}%
\begingroup\makeatletter\ifx\SetFigFont\undefined%
\gdef\SetFigFont#1#2#3#4#5{%
  \reset@font\fontsize{#1}{#2pt}%
  \fontfamily{#3}\fontseries{#4}\fontshape{#5}%
  \selectfont}%
\fi\endgroup%
\begin{picture}(3800,1638)(213,-1051)
\put(880,-988){\makebox(0,0)[lb]{\smash{{\SetFigFont{12}{14.4}{\rmdefault}{\mddefault}{\updefault}(a)}}}}
\put(2878,-971){\makebox(0,0)[lb]{\smash{{\SetFigFont{12}{14.4}{\rmdefault}{\mddefault}{\updefault}(b)}}}}
\put(835,486){\makebox(0,0)[lb]{\smash{{\SetFigFont{12}{14.4}{\rmdefault}{\mddefault}{\updefault}$a$}}}}
\put(1521,198){\makebox(0,0)[lb]{\smash{{\SetFigFont{12}{14.4}{\rmdefault}{\mddefault}{\updefault}$b$}}}}
\put(1515,-412){\makebox(0,0)[lb]{\smash{{\SetFigFont{12}{14.4}{\rmdefault}{\mddefault}{\updefault}$c$}}}}
\put(1014,-694){\makebox(0,0)[lb]{\smash{{\SetFigFont{12}{14.4}{\rmdefault}{\mddefault}{\updefault}$d$}}}}
\put(283,-495){\makebox(0,0)[lb]{\smash{{\SetFigFont{12}{14.4}{\rmdefault}{\mddefault}{\updefault}$e$}}}}
\put(213,133){\makebox(0,0)[lb]{\smash{{\SetFigFont{12}{14.4}{\rmdefault}{\mddefault}{\updefault}$f$}}}}
\put(2291,262){\makebox(0,0)[lb]{\smash{{\SetFigFont{12}{14.4}{\rmdefault}{\mddefault}{\updefault}$a$}}}}
\put(3439,390){\makebox(0,0)[lb]{\smash{{\SetFigFont{12}{14.4}{\rmdefault}{\mddefault}{\updefault}$b$}}}}
\put(2259,-579){\makebox(0,0)[lb]{\smash{{\SetFigFont{12}{14.4}{\rmdefault}{\mddefault}{\updefault}$c$}}}}
\put(3875,-277){\makebox(0,0)[lb]{\smash{{\SetFigFont{12}{14.4}{\rmdefault}{\mddefault}{\updefault}$d$}}}}
\put(2862,473){\makebox(0,0)[lb]{\smash{{\SetFigFont{12}{14.4}{\rmdefault}{\mddefault}{\updefault}$e$}}}}
\put(3292,-726){\makebox(0,0)[lb]{\smash{{\SetFigFont{12}{14.4}{\rmdefault}{\mddefault}{\updefault}$f$}}}}
\end{picture}%

%% file: g1.pstex_t
\begin{picture}(0,0)%
\includegraphics{g1.pstex}%
\end{picture}%
\setlength{\unitlength}{3947sp}%
\begingroup\makeatletter\ifx\SetFigFont\undefined%
\gdef\SetFigFont#1#2#3#4#5{%
  \reset@font\fontsize{#1}{#2pt}%
  \fontfamily{#3}\fontseries{#4}\fontshape{#5}%
  \selectfont}%
\fi\endgroup%
\begin{picture}(2805,2021)(1213,-1685)
\put(3253,-1624){\makebox(0,0)[lb]{\smash{{\SetFigFont{12}{14.4}{\rmdefault}{\mddefault}{\updefault}$g$}}}}
\put(2374,-1566){\makebox(0,0)[lb]{\smash{{\SetFigFont{12}{14.4}{\rmdefault}{\mddefault}{\updefault}$e$}}}}
\put(1213,-508){\makebox(0,0)[lb]{\smash{{\SetFigFont{12}{14.4}{\rmdefault}{\mddefault}{\updefault}$j$}}}}
\put(1989,-65){\makebox(0,0)[lb]{\smash{{\SetFigFont{12}{14.4}{\rmdefault}{\mddefault}{\updefault}$a$}}}}
\put(1842,-777){\makebox(0,0)[lb]{\smash{{\SetFigFont{12}{14.4}{\rmdefault}{\mddefault}{\updefault}$c$}}}}
\put(1361,-1445){\makebox(0,0)[lb]{\smash{{\SetFigFont{12}{14.4}{\rmdefault}{\mddefault}{\updefault}$f$}}}}
\put(2997,-822){\makebox(0,0)[lb]{\smash{{\SetFigFont{12}{14.4}{\rmdefault}{\mddefault}{\updefault}$d$}}}}
\put(2804,-72){\makebox(0,0)[lb]{\smash{{\SetFigFont{12}{14.4}{\rmdefault}{\mddefault}{\updefault}$b$}}}}
\put(3773,172){\makebox(0,0)[lb]{\smash{{\SetFigFont{12}{14.4}{\rmdefault}{\mddefault}{\updefault}$i$}}}}
\put(3593,-521){\makebox(0,0)[lb]{\smash{{\SetFigFont{12}{14.4}{\rmdefault}{\mddefault}{\updefault}$h$}}}}
\end{picture}%

%% file: d1.pstex_t
\begin{picture}(0,0)%
\includegraphics{d1.pstex}%
\end{picture}%
\setlength{\unitlength}{3947sp}%
\begingroup\makeatletter\ifx\SetFigFont\undefined%
\gdef\SetFigFont#1#2#3#4#5{%
  \reset@font\fontsize{#1}{#2pt}%
  \fontfamily{#3}\fontseries{#4}\fontshape{#5}%
  \selectfont}%
\fi\endgroup%
\begin{picture}(3795,2742)(220,-2047)
\put(2269,-909){\makebox(0,0)[lb]{\smash{{\SetFigFont{12}{14.4}{\rmdefault}{\mddefault}{\updefault}$a$}}}}
\put(2141,-839){\makebox(0,0)[lb]{\smash{{\SetFigFont{12}{14.4}{\rmdefault}{\mddefault}{\updefault}$j$}}}}
\put(1984,407){\makebox(0,0)[lb]{\smash{{\SetFigFont{12}{14.4}{\rmdefault}{\mddefault}{\updefault}$a$}}}}
\put(2199,419){\makebox(0,0)[lb]{\smash{{\SetFigFont{12}{14.4}{\rmdefault}{\mddefault}{\updefault}$b$}}}}
\put(1966,273){\makebox(0,0)[lb]{\smash{{\SetFigFont{12}{14.4}{\rmdefault}{\mddefault}{\updefault}$c$}}}}
\put(2205,273){\makebox(0,0)[lb]{\smash{{\SetFigFont{12}{14.4}{\rmdefault}{\mddefault}{\updefault}$d$}}}}
\put(1348,-718){\makebox(0,0)[lb]{\smash{{\SetFigFont{12}{14.4}{\rmdefault}{\mddefault}{\updefault}$d$}}}}
\put(3062,-566){\makebox(0,0)[lb]{\smash{{\SetFigFont{12}{14.4}{\rmdefault}{\mddefault}{\updefault}$b$}}}}
\put(2975,-724){\makebox(0,0)[lb]{\smash{{\SetFigFont{12}{14.4}{\rmdefault}{\mddefault}{\updefault}$h$}}}}
\put(2841,-566){\makebox(0,0)[lb]{\smash{{\SetFigFont{12}{14.4}{\rmdefault}{\mddefault}{\updefault}$d$}}}}
\put(3733,-1371){\makebox(0,0)[lb]{\smash{{\SetFigFont{12}{14.4}{\rmdefault}{\mddefault}{\updefault}$h$}}}}
\put(584,-1359){\makebox(0,0)[lb]{\smash{{\SetFigFont{12}{14.4}{\rmdefault}{\mddefault}{\updefault}$f$}}}}
\put(427,-1494){\makebox(0,0)[lb]{\smash{{\SetFigFont{12}{14.4}{\rmdefault}{\mddefault}{\updefault}$e$}}}}
\put(1366,-1768){\makebox(0,0)[lb]{\smash{{\SetFigFont{12}{14.4}{\rmdefault}{\mddefault}{\updefault}$g$}}}}
\put(1261,-1919){\makebox(0,0)[lb]{\smash{{\SetFigFont{12}{14.4}{\rmdefault}{\mddefault}{\updefault}$d$}}}}
\put(1342,-561){\makebox(0,0)[lb]{\smash{{\SetFigFont{12}{14.4}{\rmdefault}{\mddefault}{\updefault}$a$}}}}
\put(1558,-712){\makebox(0,0)[lb]{\smash{{\SetFigFont{12}{14.4}{\rmdefault}{\mddefault}{\updefault}$e$}}}}
\put(1546,-555){\makebox(0,0)[lb]{\smash{{\SetFigFont{12}{14.4}{\rmdefault}{\mddefault}{\updefault}$c$}}}}
\put(380,-1342){\makebox(0,0)[lb]{\smash{{\SetFigFont{12}{14.4}{\rmdefault}{\mddefault}{\updefault}$c$}}}}
\put(1196,-1756){\makebox(0,0)[lb]{\smash{{\SetFigFont{12}{14.4}{\rmdefault}{\mddefault}{\updefault}$e$}}}}
\put(3546,-1359){\makebox(0,0)[lb]{\smash{{\SetFigFont{12}{14.4}{\rmdefault}{\mddefault}{\updefault}$b$}}}}
\put(3622,-1534){\makebox(0,0)[lb]{\smash{{\SetFigFont{12}{14.4}{\rmdefault}{\mddefault}{\updefault}$i$}}}}
\end{picture}%

%% file: tree_sep.pstex_t
\begin{picture}(0,0)%
\includegraphics{tree_sep.pstex}%
\end{picture}%
\setlength{\unitlength}{3947sp}%
\begingroup\makeatletter\ifx\SetFigFont\undefined%
\gdef\SetFigFont#1#2#3#4#5{%
  \reset@font\fontsize{#1}{#2pt}%
  \fontfamily{#3}\fontseries{#4}\fontshape{#5}%
  \selectfont}%
\fi\endgroup%
\begin{picture}(2494,2614)(131,-1782)
\put(1297,-1023){\makebox(0,0)[lb]{\smash{{\SetFigFont{12}{14.4}{\rmdefault}{\mddefault}{\updefault}$y$}}}}
\put(443,-883){\makebox(0,0)[lb]{\smash{{\SetFigFont{12}{14.4}{\rmdefault}{\mddefault}{\updefault}$x$}}}}
\put(420,-1218){\makebox(0,0)[lb]{\smash{{\SetFigFont{12}{14.4}{\rmdefault}{\mddefault}{\updefault}$i$}}}}
\put(2205,-1086){\makebox(0,0)[lb]{\smash{{\SetFigFont{12}{14.4}{\rmdefault}{\mddefault}{\updefault}$p$}}}}
\put(2306,-814){\makebox(0,0)[lb]{\smash{{\SetFigFont{12}{14.4}{\rmdefault}{\mddefault}{\updefault}$y$}}}}
\put(2229,-682){\makebox(0,0)[lb]{\smash{{\SetFigFont{12}{14.4}{\rmdefault}{\mddefault}{\updefault}$x$}}}}
\put(521,668){\makebox(0,0)[lb]{\smash{{\SetFigFont{12}{14.4}{\rmdefault}{\mddefault}{\updefault}$J$}}}}
\put(1251,-1396){\makebox(0,0)[lb]{\smash{{\SetFigFont{12}{14.4}{\rmdefault}{\mddefault}{\updefault}$k$}}}}
\put(1564,339){\makebox(0,0)[lb]{\smash{{\SetFigFont{12}{14.4}{\rmdefault}{\mddefault}{\updefault}$q$}}}}
\put(1115,-1721){\makebox(0,0)[lb]{\smash{{\SetFigFont{12}{14.4}{\rmdefault}{\mddefault}{\updefault}$C_{j}$}}}}
\put(147,-1546){\makebox(0,0)[lb]{\smash{{\SetFigFont{12}{14.4}{\rmdefault}{\mddefault}{\updefault}$C_{i}$}}}}
\end{picture}%

%% file: baker_subs.pstex_t
\begin{picture}(0,0)%
\includegraphics{baker_subs.pstex}%
\end{picture}%
\setlength{\unitlength}{3947sp}%
\begingroup\makeatletter\ifx\SetFigFont\undefined%
\gdef\SetFigFont#1#2#3#4#5{%
  \reset@font\fontsize{#1}{#2pt}%
  \fontfamily{#3}\fontseries{#4}\fontshape{#5}%
  \selectfont}%
\fi\endgroup%
\begin{picture}(6037,3201)(634,-3286)
\put(5208,-3225){\makebox(0,0)[lb]{\smash{{\SetFigFont{12}{14.4}{\rmdefault}{\mddefault}{\updefault}(b)}}}}
\put(1871,-3190){\makebox(0,0)[lb]{\smash{{\SetFigFont{12}{14.4}{\rmdefault}{\mddefault}{\updefault}(a)}}}}
\end{picture}%

%% file: baker1.pstex_t
\begin{picture}(0,0)%
\includegraphics{baker1.pstex}%
\end{picture}%
\setlength{\unitlength}{3947sp}%
\begingroup\makeatletter\ifx\SetFigFont\undefined%
\gdef\SetFigFont#1#2#3#4#5{%
  \reset@font\fontsize{#1}{#2pt}%
  \fontfamily{#3}\fontseries{#4}\fontshape{#5}%
  \selectfont}%
\fi\endgroup%
\begin{picture}(2373,2033)(138,-1311)
\end{picture}%

%% file: baker2.pstex_t
\begin{picture}(0,0)%
\includegraphics{baker2.pstex}%
\end{picture}%
\setlength{\unitlength}{3947sp}%
\begingroup\makeatletter\ifx\SetFigFont\undefined%
\gdef\SetFigFont#1#2#3#4#5{%
  \reset@font\fontsize{#1}{#2pt}%
  \fontfamily{#3}\fontseries{#4}\fontshape{#5}%
  \selectfont}%
\fi\endgroup%
\begin{picture}(4920,3638)(174,-2958)
\put(3144,-1374){\makebox(0,0)[lb]{\smash{{\SetFigFont{12}{14.4}{\rmdefault}{\mddefault}{\updefault}3:2}}}}
\put(3548,-919){\makebox(0,0)[lb]{\smash{{\SetFigFont{12}{14.4}{\rmdefault}{\mddefault}{\updefault}3:0}}}}
\put(2920,-495){\makebox(0,0)[lb]{\smash{{\SetFigFont{12}{14.4}{\rmdefault}{\mddefault}{\updefault}3:1}}}}
\put(2625,-2278){\makebox(0,0)[lb]{\smash{{\SetFigFont{12}{14.4}{\rmdefault}{\mddefault}{\updefault}2:3}}}}
\put(4658,-117){\makebox(0,0)[lb]{\smash{{\SetFigFont{12}{14.4}{\rmdefault}{\mddefault}{\updefault}1:0}}}}
\put(4074,-2009){\makebox(0,0)[lb]{\smash{{\SetFigFont{12}{14.4}{\rmdefault}{\mddefault}{\updefault}2:4}}}}
\put(1040,-1829){\makebox(0,0)[lb]{\smash{{\SetFigFont{12}{14.4}{\rmdefault}{\mddefault}{\updefault}2:7}}}}
\put(751,-1194){\makebox(0,0)[lb]{\smash{{\SetFigFont{12}{14.4}{\rmdefault}{\mddefault}{\updefault}2:6}}}}
\put(1175,-726){\makebox(0,0)[lb]{\smash{{\SetFigFont{12}{14.4}{\rmdefault}{\mddefault}{\updefault}2:5}}}}
\put(2054,-1002){\makebox(0,0)[lb]{\smash{{\SetFigFont{12}{14.4}{\rmdefault}{\mddefault}{\updefault}2:2}}}}
\put(2714, -8){\makebox(0,0)[lb]{\smash{{\SetFigFont{12}{14.4}{\rmdefault}{\mddefault}{\updefault}2:1}}}}
\put(2785,480){\makebox(0,0)[lb]{\smash{{\SetFigFont{12}{14.4}{\rmdefault}{\mddefault}{\updefault}1:1}}}}
\put(1329,294){\makebox(0,0)[lb]{\smash{{\SetFigFont{12}{14.4}{\rmdefault}{\mddefault}{\updefault}1:2}}}}
\put(174, -1){\makebox(0,0)[lb]{\smash{{\SetFigFont{12}{14.4}{\rmdefault}{\mddefault}{\updefault}1:3}}}}
\put(264,-2413){\makebox(0,0)[lb]{\smash{{\SetFigFont{12}{14.4}{\rmdefault}{\mddefault}{\updefault}1:4}}}}
\put(1534,-2715){\makebox(0,0)[lb]{\smash{{\SetFigFont{12}{14.4}{\rmdefault}{\mddefault}{\updefault}1:5}}}}
\put(2663,-2958){\makebox(0,0)[lb]{\smash{{\SetFigFont{12}{14.4}{\rmdefault}{\mddefault}{\updefault}1:6}}}}
\put(4209,-2567){\makebox(0,0)[lb]{\smash{{\SetFigFont{12}{14.4}{\rmdefault}{\mddefault}{\updefault}1:7}}}}
\put(4876,-1432){\makebox(0,0)[lb]{\smash{{\SetFigFont{12}{14.4}{\rmdefault}{\mddefault}{\updefault}1:8}}}}
\put(3933,-457){\makebox(0,0)[lb]{\smash{{\SetFigFont{12}{14.4}{\rmdefault}{\mddefault}{\updefault}2:0}}}}
\end{picture}%

%% file: baker3.pstex_t
\begin{picture}(0,0)%
\includegraphics{baker3.pstex}%
\end{picture}%
\setlength{\unitlength}{3947sp}%
\begingroup\makeatletter\ifx\SetFigFont\undefined%
\gdef\SetFigFont#1#2#3#4#5{%
  \reset@font\fontsize{#1}{#2pt}%
  \fontfamily{#3}\fontseries{#4}\fontshape{#5}%
  \selectfont}%
\fi\endgroup%
\begin{picture}(6057,4229)(174,-3549)
\put(4685,-1510){\makebox(0,0)[lb]{\smash{{\SetFigFont{12}{14.4}{\rmdefault}{\mddefault}{\updefault}3:0}}}}
\put(3393,-817){\makebox(0,0)[lb]{\smash{{\SetFigFont{12}{14.4}{\rmdefault}{\mddefault}{\updefault}3:1}}}}
\put(5131,-484){\makebox(0,0)[lb]{\smash{{\SetFigFont{12}{14.4}{\rmdefault}{\mddefault}{\updefault}1:0}}}}
\put(4281,-1965){\makebox(0,0)[lb]{\smash{{\SetFigFont{12}{14.4}{\rmdefault}{\mddefault}{\updefault}3:2}}}}
\put(3762,-2869){\makebox(0,0)[lb]{\smash{{\SetFigFont{12}{14.4}{\rmdefault}{\mddefault}{\updefault}2:3}}}}
\put(5211,-2600){\makebox(0,0)[lb]{\smash{{\SetFigFont{12}{14.4}{\rmdefault}{\mddefault}{\updefault}2:4}}}}
\put(3800,-3549){\makebox(0,0)[lb]{\smash{{\SetFigFont{12}{14.4}{\rmdefault}{\mddefault}{\updefault}1:6}}}}
\put(5346,-3158){\makebox(0,0)[lb]{\smash{{\SetFigFont{12}{14.4}{\rmdefault}{\mddefault}{\updefault}1:7}}}}
\put(6013,-2023){\makebox(0,0)[lb]{\smash{{\SetFigFont{12}{14.4}{\rmdefault}{\mddefault}{\updefault}1:8}}}}
\put(2708,-3277){\makebox(0,0)[lb]{\smash{{\SetFigFont{12}{14.4}{\rmdefault}{\mddefault}{\updefault}1:5}}}}
\put(2638,-2249){\makebox(0,0)[lb]{\smash{{\SetFigFont{12}{14.4}{\rmdefault}{\mddefault}{\updefault}2:3}}}}
\put(1534,-2715){\makebox(0,0)[lb]{\smash{{\SetFigFont{12}{14.4}{\rmdefault}{\mddefault}{\updefault}1:5}}}}
\put(264,-2413){\makebox(0,0)[lb]{\smash{{\SetFigFont{12}{14.4}{\rmdefault}{\mddefault}{\updefault}1:4}}}}
\put(2920,-450){\makebox(0,0)[lb]{\smash{{\SetFigFont{12}{14.4}{\rmdefault}{\mddefault}{\updefault}3:1}}}}
\put(4021,-1286){\makebox(0,0)[lb]{\smash{{\SetFigFont{12}{14.4}{\rmdefault}{\mddefault}{\updefault}3:0}}}}
\put(5805,-723){\makebox(0,0)[lb]{\smash{{\SetFigFont{12}{14.4}{\rmdefault}{\mddefault}{\updefault}1:0}}}}
\put(174, -1){\makebox(0,0)[lb]{\smash{{\SetFigFont{12}{14.4}{\rmdefault}{\mddefault}{\updefault}1:3}}}}
\put(3144,-1374){\makebox(0,0)[lb]{\smash{{\SetFigFont{12}{14.4}{\rmdefault}{\mddefault}{\updefault}3:2}}}}
\put(4658,-117){\makebox(0,0)[lb]{\smash{{\SetFigFont{12}{14.4}{\rmdefault}{\mddefault}{\updefault}1:0}}}}
\put(1040,-1829){\makebox(0,0)[lb]{\smash{{\SetFigFont{12}{14.4}{\rmdefault}{\mddefault}{\updefault}2:7}}}}
\put(751,-1194){\makebox(0,0)[lb]{\smash{{\SetFigFont{12}{14.4}{\rmdefault}{\mddefault}{\updefault}2:6}}}}
\put(1175,-726){\makebox(0,0)[lb]{\smash{{\SetFigFont{12}{14.4}{\rmdefault}{\mddefault}{\updefault}2:5}}}}
\put(2054,-1002){\makebox(0,0)[lb]{\smash{{\SetFigFont{12}{14.4}{\rmdefault}{\mddefault}{\updefault}2:2}}}}
\put(2714, -8){\makebox(0,0)[lb]{\smash{{\SetFigFont{12}{14.4}{\rmdefault}{\mddefault}{\updefault}2:1}}}}
\put(2785,480){\makebox(0,0)[lb]{\smash{{\SetFigFont{12}{14.4}{\rmdefault}{\mddefault}{\updefault}1:1}}}}
\put(1329,294){\makebox(0,0)[lb]{\smash{{\SetFigFont{12}{14.4}{\rmdefault}{\mddefault}{\updefault}1:2}}}}
\put(3933,-457){\makebox(0,0)[lb]{\smash{{\SetFigFont{12}{14.4}{\rmdefault}{\mddefault}{\updefault}2:0}}}}
\put(5070,-1048){\makebox(0,0)[lb]{\smash{{\SetFigFont{12}{14.4}{\rmdefault}{\mddefault}{\updefault}2:0}}}}
\put(4406,-824){\makebox(0,0)[lb]{\smash{{\SetFigFont{12}{14.4}{\rmdefault}{\mddefault}{\updefault}2:0}}}}
\end{picture}%

%% file: baker4.pstex_t
\begin{picture}(0,0)%
\includegraphics{baker4.pstex}%
\end{picture}%
\setlength{\unitlength}{3947sp}%
\begingroup\makeatletter\ifx\SetFigFont\undefined%
\gdef\SetFigFont#1#2#3#4#5{%
  \reset@font\fontsize{#1}{#2pt}%
  \fontfamily{#3}\fontseries{#4}\fontshape{#5}%
  \selectfont}%
\fi\endgroup%
\begin{picture}(5798,3526)(965,-3212)
\put(6545,-809){\makebox(0,0)[lb]{\smash{{\SetFigFont{12}{14.4}{\rmdefault}{\mddefault}{\updefault}1:0}}}}
\put(4601,-700){\makebox(0,0)[lb]{\smash{{\SetFigFont{12}{14.4}{\rmdefault}{\mddefault}{\updefault}2:1}}}}
\put(4672,-212){\makebox(0,0)[lb]{\smash{{\SetFigFont{12}{14.4}{\rmdefault}{\mddefault}{\updefault}1:1}}}}
\put(965,-498){\makebox(0,0)[lb]{\smash{{\SetFigFont{12}{14.4}{\rmdefault}{\mddefault}{\updefault}1:3}}}}
\put(3385,-2776){\makebox(0,0)[lb]{\smash{{\SetFigFont{12}{14.4}{\rmdefault}{\mddefault}{\updefault}2:3}}}}
\put(4098,-374){\makebox(0,0)[lb]{\smash{{\SetFigFont{12}{14.4}{\rmdefault}{\mddefault}{\updefault}2:1}}}}
\put(4169,114){\makebox(0,0)[lb]{\smash{{\SetFigFont{12}{14.4}{\rmdefault}{\mddefault}{\updefault}1:1}}}}
\put(3438,-1368){\makebox(0,0)[lb]{\smash{{\SetFigFont{12}{14.4}{\rmdefault}{\mddefault}{\updefault}2:2}}}}
\put(2713,-72){\makebox(0,0)[lb]{\smash{{\SetFigFont{12}{14.4}{\rmdefault}{\mddefault}{\updefault}1:2}}}}
\put(2845,-1499){\makebox(0,0)[lb]{\smash{{\SetFigFont{12}{14.4}{\rmdefault}{\mddefault}{\updefault}2:2}}}}
\put(2120,-203){\makebox(0,0)[lb]{\smash{{\SetFigFont{12}{14.4}{\rmdefault}{\mddefault}{\updefault}1:2}}}}
\put(2325,-3212){\makebox(0,0)[lb]{\smash{{\SetFigFont{12}{14.4}{\rmdefault}{\mddefault}{\updefault}1:5}}}}
\put(1831,-2326){\makebox(0,0)[lb]{\smash{{\SetFigFont{12}{14.4}{\rmdefault}{\mddefault}{\updefault}2:7}}}}
\put(1542,-1691){\makebox(0,0)[lb]{\smash{{\SetFigFont{12}{14.4}{\rmdefault}{\mddefault}{\updefault}2:6}}}}
\put(1966,-1223){\makebox(0,0)[lb]{\smash{{\SetFigFont{12}{14.4}{\rmdefault}{\mddefault}{\updefault}2:5}}}}
\put(1055,-2910){\makebox(0,0)[lb]{\smash{{\SetFigFont{12}{14.4}{\rmdefault}{\mddefault}{\updefault}1:4}}}}
\put(5861,-1189){\makebox(0,0)[lb]{\smash{{\SetFigFont{12}{14.4}{\rmdefault}{\mddefault}{\updefault}2:0}}}}
\end{picture}%

%% file: baker5.pstex_t
\begin{picture}(0,0)%
\includegraphics{baker5.pstex}%
\end{picture}%
\setlength{\unitlength}{3947sp}%
\begingroup\makeatletter\ifx\SetFigFont\undefined%
\gdef\SetFigFont#1#2#3#4#5{%
  \reset@font\fontsize{#1}{#2pt}%
  \fontfamily{#3}\fontseries{#4}\fontshape{#5}%
  \selectfont}%
\fi\endgroup%
\begin{picture}(3058,2986)(85,-2185)
\put(2431,-970){\makebox(0,0)[lb]{\smash{{\SetFigFont{12}{14.4}{\rmdefault}{\mddefault}{\updefault}2:7}}}}
\put(1655,-1554){\makebox(0,0)[lb]{\smash{{\SetFigFont{12}{14.4}{\rmdefault}{\mddefault}{\updefault}1:4}}}}
\put(2566,133){\makebox(0,0)[lb]{\smash{{\SetFigFont{12}{14.4}{\rmdefault}{\mddefault}{\updefault}2:5}}}}
\put(2925,-1856){\makebox(0,0)[lb]{\smash{{\SetFigFont{12}{14.4}{\rmdefault}{\mddefault}{\updefault}1:5}}}}
\put(1846,601){\makebox(0,0)[lb]{\smash{{\SetFigFont{12}{14.4}{\rmdefault}{\mddefault}{\updefault}1:2}}}}
\put(1268,-887){\makebox(0,0)[lb]{\smash{{\SetFigFont{12}{14.4}{\rmdefault}{\mddefault}{\updefault}2:6}}}}
\put(1692,-419){\makebox(0,0)[lb]{\smash{{\SetFigFont{12}{14.4}{\rmdefault}{\mddefault}{\updefault}2:5}}}}
\put(691,306){\makebox(0,0)[lb]{\smash{{\SetFigFont{12}{14.4}{\rmdefault}{\mddefault}{\updefault}1:3}}}}
\put(662,-966){\makebox(0,0)[lb]{\smash{{\SetFigFont{12}{14.4}{\rmdefault}{\mddefault}{\updefault}2:6}}}}
\put( 85,227){\makebox(0,0)[lb]{\smash{{\SetFigFont{12}{14.4}{\rmdefault}{\mddefault}{\updefault}1:3}}}}
\put(951,-1601){\makebox(0,0)[lb]{\smash{{\SetFigFont{12}{14.4}{\rmdefault}{\mddefault}{\updefault}2:7}}}}
\put(175,-2185){\makebox(0,0)[lb]{\smash{{\SetFigFont{12}{14.4}{\rmdefault}{\mddefault}{\updefault}1:4}}}}
\end{picture}%

%% file: d2.pstex_t
\begin{picture}(0,0)%
\includegraphics{d2.pstex}%
\end{picture}%
\setlength{\unitlength}{3947sp}%
\begingroup\makeatletter\ifx\SetFigFont\undefined%
\gdef\SetFigFont#1#2#3#4#5{%
  \reset@font\fontsize{#1}{#2pt}%
  \fontfamily{#3}\fontseries{#4}\fontshape{#5}%
  \selectfont}%
\fi\endgroup%
\begin{picture}(4231,2742)(220,-2047)
\put(2269,-909){\makebox(0,0)[lb]{\smash{{\SetFigFont{12}{14.4}{\rmdefault}{\mddefault}{\updefault}$a$}}}}
\put(2141,-839){\makebox(0,0)[lb]{\smash{{\SetFigFont{12}{14.4}{\rmdefault}{\mddefault}{\updefault}$j$}}}}
\put(1984,407){\makebox(0,0)[lb]{\smash{{\SetFigFont{12}{14.4}{\rmdefault}{\mddefault}{\updefault}$a$}}}}
\put(2199,419){\makebox(0,0)[lb]{\smash{{\SetFigFont{12}{14.4}{\rmdefault}{\mddefault}{\updefault}$b$}}}}
\put(1966,273){\makebox(0,0)[lb]{\smash{{\SetFigFont{12}{14.4}{\rmdefault}{\mddefault}{\updefault}$c$}}}}
\put(2205,273){\makebox(0,0)[lb]{\smash{{\SetFigFont{12}{14.4}{\rmdefault}{\mddefault}{\updefault}$d$}}}}
\put(1348,-718){\makebox(0,0)[lb]{\smash{{\SetFigFont{12}{14.4}{\rmdefault}{\mddefault}{\updefault}$d$}}}}
\put(3733,-1371){\makebox(0,0)[lb]{\smash{{\SetFigFont{12}{14.4}{\rmdefault}{\mddefault}{\updefault}$h$}}}}
\put(584,-1359){\makebox(0,0)[lb]{\smash{{\SetFigFont{12}{14.4}{\rmdefault}{\mddefault}{\updefault}$f$}}}}
\put(427,-1494){\makebox(0,0)[lb]{\smash{{\SetFigFont{12}{14.4}{\rmdefault}{\mddefault}{\updefault}$e$}}}}
\put(1366,-1768){\makebox(0,0)[lb]{\smash{{\SetFigFont{12}{14.4}{\rmdefault}{\mddefault}{\updefault}$g$}}}}
\put(1261,-1919){\makebox(0,0)[lb]{\smash{{\SetFigFont{12}{14.4}{\rmdefault}{\mddefault}{\updefault}$d$}}}}
\put(1342,-561){\makebox(0,0)[lb]{\smash{{\SetFigFont{12}{14.4}{\rmdefault}{\mddefault}{\updefault}$a$}}}}
\put(1558,-712){\makebox(0,0)[lb]{\smash{{\SetFigFont{12}{14.4}{\rmdefault}{\mddefault}{\updefault}$e$}}}}
\put(1546,-555){\makebox(0,0)[lb]{\smash{{\SetFigFont{12}{14.4}{\rmdefault}{\mddefault}{\updefault}$c$}}}}
\put(380,-1342){\makebox(0,0)[lb]{\smash{{\SetFigFont{12}{14.4}{\rmdefault}{\mddefault}{\updefault}$c$}}}}
\put(1196,-1756){\makebox(0,0)[lb]{\smash{{\SetFigFont{12}{14.4}{\rmdefault}{\mddefault}{\updefault}$e$}}}}
\put(3062,-566){\makebox(0,0)[lb]{\smash{{\SetFigFont{12}{14.4}{\rmdefault}{\mddefault}{\updefault}$b$}}}}
\put(3947,-513){\makebox(0,0)[lb]{\smash{{\SetFigFont{12}{14.4}{\rmdefault}{\mddefault}{\updefault}$d$}}}}
\put(3745,-1517){\makebox(0,0)[lb]{\smash{{\SetFigFont{12}{14.4}{\rmdefault}{\mddefault}{\updefault}$i$}}}}
\put(3529,-1546){\makebox(0,0)[lb]{\smash{{\SetFigFont{12}{14.4}{\rmdefault}{\mddefault}{\updefault}$d$}}}}
\put(3541,-1377){\makebox(0,0)[lb]{\smash{{\SetFigFont{12}{14.4}{\rmdefault}{\mddefault}{\updefault}$b$}}}}
\put(3045,-718){\makebox(0,0)[lb]{\smash{{\SetFigFont{12}{14.4}{\rmdefault}{\mddefault}{\updefault}$h$}}}}
\put(2876,-747){\makebox(0,0)[lb]{\smash{{\SetFigFont{12}{14.4}{\rmdefault}{\mddefault}{\updefault}$i$}}}}
\put(2870,-549){\makebox(0,0)[lb]{\smash{{\SetFigFont{12}{14.4}{\rmdefault}{\mddefault}{\updefault}$d$}}}}
\put(4159,-508){\makebox(0,0)[lb]{\smash{{\SetFigFont{12}{14.4}{\rmdefault}{\mddefault}{\updefault}$b$}}}}
\put(4170,-642){\makebox(0,0)[lb]{\smash{{\SetFigFont{12}{14.4}{\rmdefault}{\mddefault}{\updefault}$h$}}}}
\put(3984,-683){\makebox(0,0)[lb]{\smash{{\SetFigFont{12}{14.4}{\rmdefault}{\mddefault}{\updefault}$i$}}}}
\end{picture}%

%% file: d4.pstex_t
\begin{picture}(0,0)%
\includegraphics{d4.pstex}%
\end{picture}%
\setlength{\unitlength}{3947sp}%
\begingroup\makeatletter\ifx\SetFigFont\undefined%
\gdef\SetFigFont#1#2#3#4#5{%
  \reset@font\fontsize{#1}{#2pt}%
  \fontfamily{#3}\fontseries{#4}\fontshape{#5}%
  \selectfont}%
\fi\endgroup%
\begin{picture}(5248,3146)(358,-2741)
\put(4671, 50){\makebox(0,0)[lb]{\smash{{\SetFigFont{12}{14.4}{\rmdefault}{\mddefault}{\updefault}$i$}}}}
\put(4491, 89){\makebox(0,0)[lb]{\smash{{\SetFigFont{12}{14.4}{\rmdefault}{\mddefault}{\updefault}$b$}}}}
\put(3151,-1618){\makebox(0,0)[lb]{\smash{{\SetFigFont{12}{14.4}{\rmdefault}{\mddefault}{\updefault}$e$}}}}
\put(3349,-1663){\makebox(0,0)[lb]{\smash{{\SetFigFont{12}{14.4}{\rmdefault}{\mddefault}{\updefault}$d$}}}}
\put(1149,-508){\makebox(0,0)[lb]{\smash{{\SetFigFont{12}{14.4}{\rmdefault}{\mddefault}{\updefault}$a$}}}}
\put(1348,-463){\makebox(0,0)[lb]{\smash{{\SetFigFont{12}{14.4}{\rmdefault}{\mddefault}{\updefault}$c$}}}}
\put(1191,-1232){\makebox(0,0)[lb]{\smash{{\SetFigFont{12}{14.4}{\rmdefault}{\mddefault}{\updefault}$d$}}}}
\put(1409,-1257){\makebox(0,0)[lb]{\smash{{\SetFigFont{12}{14.4}{\rmdefault}{\mddefault}{\updefault}$a$}}}}
\put(1185,-1936){\makebox(0,0)[lb]{\smash{{\SetFigFont{12}{14.4}{\rmdefault}{\mddefault}{\updefault}$a$}}}}
\put(1371,-1898){\makebox(0,0)[lb]{\smash{{\SetFigFont{12}{14.4}{\rmdefault}{\mddefault}{\updefault}$e$}}}}
\put(701,-2383){\makebox(0,0)[lb]{\smash{{\SetFigFont{12}{14.4}{\rmdefault}{\mddefault}{\updefault}$f$}}}}
\put(534,-2460){\makebox(0,0)[lb]{\smash{{\SetFigFont{12}{14.4}{\rmdefault}{\mddefault}{\updefault}$c$}}}}
\put(4139,-1277){\makebox(0,0)[lb]{\smash{{\SetFigFont{12}{14.4}{\rmdefault}{\mddefault}{\updefault}$d$}}}}
\put(4318,-1341){\makebox(0,0)[lb]{\smash{{\SetFigFont{12}{14.4}{\rmdefault}{\mddefault}{\updefault}$h$}}}}
\put(630,-132){\makebox(0,0)[lb]{\smash{{\SetFigFont{12}{14.4}{\rmdefault}{\mddefault}{\updefault}$j$}}}}
\put(437,-228){\makebox(0,0)[lb]{\smash{{\SetFigFont{12}{14.4}{\rmdefault}{\mddefault}{\updefault}$a$}}}}
\put(2224,-2236){\makebox(0,0)[lb]{\smash{{\SetFigFont{12}{14.4}{\rmdefault}{\mddefault}{\updefault}$c$}}}}
\put(2404,-2191){\makebox(0,0)[lb]{\smash{{\SetFigFont{12}{14.4}{\rmdefault}{\mddefault}{\updefault}$e$}}}}
\put(2381,-1424){\makebox(0,0)[lb]{\smash{{\SetFigFont{12}{14.4}{\rmdefault}{\mddefault}{\updefault}$d$}}}}
\put(2201,-1398){\makebox(0,0)[lb]{\smash{{\SetFigFont{12}{14.4}{\rmdefault}{\mddefault}{\updefault}$c$}}}}
\put(3290,-2594){\makebox(0,0)[lb]{\smash{{\SetFigFont{12}{14.4}{\rmdefault}{\mddefault}{\updefault}$g$}}}}
\put(3161,-2549){\makebox(0,0)[lb]{\smash{{\SetFigFont{12}{14.4}{\rmdefault}{\mddefault}{\updefault}$e$}}}}
\put(4065,-2034){\makebox(0,0)[lb]{\smash{{\SetFigFont{12}{14.4}{\rmdefault}{\mddefault}{\updefault}$g$}}}}
\put(4283,-1989){\makebox(0,0)[lb]{\smash{{\SetFigFont{12}{14.4}{\rmdefault}{\mddefault}{\updefault}$d$}}}}
\put(3618,-493){\makebox(0,0)[lb]{\smash{{\SetFigFont{12}{14.4}{\rmdefault}{\mddefault}{\updefault}$d$}}}}
\put(3836,-493){\makebox(0,0)[lb]{\smash{{\SetFigFont{12}{14.4}{\rmdefault}{\mddefault}{\updefault}$b$}}}}
\put(5414,-731){\makebox(0,0)[lb]{\smash{{\SetFigFont{12}{14.4}{\rmdefault}{\mddefault}{\updefault}$i$}}}}
\put(5209,-725){\makebox(0,0)[lb]{\smash{{\SetFigFont{12}{14.4}{\rmdefault}{\mddefault}{\updefault}$h$}}}}
\put(4661,-550){\makebox(0,0)[lb]{\smash{{\SetFigFont{12}{14.4}{\rmdefault}{\mddefault}{\updefault}$h$}}}}
\put(4520,-627){\makebox(0,0)[lb]{\smash{{\SetFigFont{12}{14.4}{\rmdefault}{\mddefault}{\updefault}$b$}}}}
\put(2818,-396){\makebox(0,0)[lb]{\smash{{\SetFigFont{12}{14.4}{\rmdefault}{\mddefault}{\updefault}$c$}}}}
\put(3029,-386){\makebox(0,0)[lb]{\smash{{\SetFigFont{12}{14.4}{\rmdefault}{\mddefault}{\updefault}$b$}}}}
\end{picture}%

%% file: circle_const.pstex_t
\begin{picture}(0,0)%
\includegraphics{circle_const.pstex}%
\end{picture}%
\setlength{\unitlength}{3947sp}%
\begingroup\makeatletter\ifx\SetFigFont\undefined%
\gdef\SetFigFont#1#2#3#4#5{%
  \reset@font\fontsize{#1}{#2pt}%
  \fontfamily{#3}\fontseries{#4}\fontshape{#5}%
  \selectfont}%
\fi\endgroup%
\begin{picture}(4230,3116)(555,-2289)
\put(2012,493){\makebox(0,0)[lb]{\smash{{\SetFigFont{12}{14.4}{\rmdefault}{\mddefault}{\updefault}$v_{0}$ $v_{1}$}}}}
\put(2492,-276){\makebox(0,0)[lb]{\smash{{\SetFigFont{12}{14.4}{\rmdefault}{\mddefault}{\updefault}$v_{0}$ $v_{2}$}}}}
\put(3480,-445){\makebox(0,0)[lb]{\smash{{\SetFigFont{12}{14.4}{\rmdefault}{\mddefault}{\updefault}$v_{1}$ $v_{2}$}}}}
\put(2710,-1123){\makebox(0,0)[lb]{\smash{{\SetFigFont{12}{14.4}{\rmdefault}{\mddefault}{\updefault}$v_{1}$ $v_{3}$}}}}
\put(3127,-2047){\makebox(0,0)[lb]{\smash{{\SetFigFont{12}{14.4}{\rmdefault}{\mddefault}{\updefault}$v_{2}$ $v_{3}$}}}}
\put(848,-1948){\makebox(0,0)[lb]{\smash{{\SetFigFont{12}{14.4}{\rmdefault}{\mddefault}{\updefault}$v_{3}$ $v_{4}$}}}}
\put(2075,-1701){\makebox(0,0)[lb]{\smash{{\SetFigFont{12}{14.4}{\rmdefault}{\mddefault}{\updefault}$v_{2}$ $v_{4}$}}}}
\put(622,-353){\makebox(0,0)[lb]{\smash{{\SetFigFont{12}{14.4}{\rmdefault}{\mddefault}{\updefault}$v_{4}$ $v_{0}$}}}}
\put(1250,-1108){\makebox(0,0)[lb]{\smash{{\SetFigFont{12}{14.4}{\rmdefault}{\mddefault}{\updefault}$v_{3}$ $v_{0}$}}}}
\put(1497,-332){\makebox(0,0)[lb]{\smash{{\SetFigFont{12}{14.4}{\rmdefault}{\mddefault}{\updefault}$v_{4}$ $v_{1}$}}}}
\put(2671,109){\makebox(0,0)[lb]{\smash{{\SetFigFont{12}{14.4}{\rmdefault}{\mddefault}{\updefault}5}}}}
\put(653, 40){\makebox(0,0)[lb]{\smash{{\SetFigFont{12}{14.4}{\rmdefault}{\mddefault}{\updefault}4}}}}
\put(676,-2281){\makebox(0,0)[lb]{\smash{{\SetFigFont{12}{14.4}{\rmdefault}{\mddefault}{\updefault}3}}}}
\put(3626,-2289){\makebox(0,0)[lb]{\smash{{\SetFigFont{12}{14.4}{\rmdefault}{\mddefault}{\updefault}2}}}}
\put(3075,-721){\makebox(0,0)[lb]{\smash{{\SetFigFont{12}{14.4}{\rmdefault}{\mddefault}{\updefault}6}}}}
\put(3998,-139){\makebox(0,0)[lb]{\smash{{\SetFigFont{12}{14.4}{\rmdefault}{\mddefault}{\updefault}1}}}}
\put(2562,699){\makebox(0,0)[lb]{\smash{{\SetFigFont{12}{14.4}{\rmdefault}{\mddefault}{\updefault}0}}}}
\put(2523,-2002){\makebox(0,0)[lb]{\smash{{\SetFigFont{12}{14.4}{\rmdefault}{\mddefault}{\updefault}7}}}}
\put(1421, -7){\makebox(0,0)[lb]{\smash{{\SetFigFont{12}{14.4}{\rmdefault}{\mddefault}{\updefault}9}}}}
\put(1033,-1358){\makebox(0,0)[lb]{\smash{{\SetFigFont{12}{14.4}{\rmdefault}{\mddefault}{\updefault}8}}}}
\end{picture}%

%% file: triangle_const.pstex_t
\begin{picture}(0,0)%
\includegraphics{triangle_const.pstex}%
\end{picture}%
\setlength{\unitlength}{3947sp}%
\begingroup\makeatletter\ifx\SetFigFont\undefined%
\gdef\SetFigFont#1#2#3#4#5{%
  \reset@font\fontsize{#1}{#2pt}%
  \fontfamily{#3}\fontseries{#4}\fontshape{#5}%
  \selectfont}%
\fi\endgroup%
\begin{picture}(2434,2527)(969,-1803)
\end{picture}%

%% file: d5.pstex_t
\begin{picture}(0,0)%
\includegraphics{d5.pstex}%
\end{picture}%
\setlength{\unitlength}{3947sp}%
\begingroup\makeatletter\ifx\SetFigFont\undefined%
\gdef\SetFigFont#1#2#3#4#5{%
  \reset@font\fontsize{#1}{#2pt}%
  \fontfamily{#3}\fontseries{#4}\fontshape{#5}%
  \selectfont}%
\fi\endgroup%
\begin{picture}(2766,2452)(1017,-1750)
\put(1316,-495){\makebox(0,0)[lb]{\smash{{\SetFigFont{12}{14.4}{\rmdefault}{\mddefault}{\updefault}$a$}}}}
\put(1476,-540){\makebox(0,0)[lb]{\smash{{\SetFigFont{12}{14.4}{\rmdefault}{\mddefault}{\updefault}$c$}}}}
\put(1367,-591){\makebox(0,0)[lb]{\smash{{\SetFigFont{12}{14.4}{\rmdefault}{\mddefault}{\updefault}$e$}}}}
\put(1566,-624){\makebox(0,0)[lb]{\smash{{\SetFigFont{12}{14.4}{\rmdefault}{\mddefault}{\updefault}$d$}}}}
\put(1265,326){\makebox(0,0)[lb]{\smash{{\SetFigFont{12}{14.4}{\rmdefault}{\mddefault}{\updefault}$a$}}}}
\put(1470,403){\makebox(0,0)[lb]{\smash{{\SetFigFont{12}{14.4}{\rmdefault}{\mddefault}{\updefault}$j$}}}}
\put(1778,-1528){\makebox(0,0)[lb]{\smash{{\SetFigFont{12}{14.4}{\rmdefault}{\mddefault}{\updefault}$e$}}}}
\put(1957,-1438){\makebox(0,0)[lb]{\smash{{\SetFigFont{12}{14.4}{\rmdefault}{\mddefault}{\updefault}$d$}}}}
\put(1919,-1579){\makebox(0,0)[lb]{\smash{{\SetFigFont{12}{14.4}{\rmdefault}{\mddefault}{\updefault}$g$}}}}
\put(2727, 82){\makebox(0,0)[lb]{\smash{{\SetFigFont{12}{14.4}{\rmdefault}{\mddefault}{\updefault}$c$}}}}
\put(2913, 44){\makebox(0,0)[lb]{\smash{{\SetFigFont{12}{14.4}{\rmdefault}{\mddefault}{\updefault}$f$}}}}
\put(2734,-117){\makebox(0,0)[lb]{\smash{{\SetFigFont{12}{14.4}{\rmdefault}{\mddefault}{\updefault}$b$}}}}
\put(2932,-117){\makebox(0,0)[lb]{\smash{{\SetFigFont{12}{14.4}{\rmdefault}{\mddefault}{\updefault}$d$}}}}
\put(3266,-983){\makebox(0,0)[lb]{\smash{{\SetFigFont{12}{14.4}{\rmdefault}{\mddefault}{\updefault}$h$}}}}
\put(3490,-1047){\makebox(0,0)[lb]{\smash{{\SetFigFont{12}{14.4}{\rmdefault}{\mddefault}{\updefault}$i$}}}}
\end{picture}%

%% file: counter1.pstex_t
\begin{picture}(0,0)%
\includegraphics{counter1.pstex}%
\end{picture}%
\setlength{\unitlength}{3947sp}%
\begingroup\makeatletter\ifx\SetFigFont\undefined%
\gdef\SetFigFont#1#2#3#4#5{%
  \reset@font\fontsize{#1}{#2pt}%
  \fontfamily{#3}\fontseries{#4}\fontshape{#5}%
  \selectfont}%
\fi\endgroup%
\begin{picture}(3815,2390)(376,-1796)
\put(1483,338){\makebox(0,0)[lb]{\smash{{\SetFigFont{12}{14.4}{\rmdefault}{\mddefault}{\updefault}$a$}}}}
\put(1990,312){\makebox(0,0)[lb]{\smash{{\SetFigFont{12}{14.4}{\rmdefault}{\mddefault}{\updefault}$b$}}}}
\put(2542,299){\makebox(0,0)[lb]{\smash{{\SetFigFont{12}{14.4}{\rmdefault}{\mddefault}{\updefault}$c$}}}}
\put(3061,338){\makebox(0,0)[lb]{\smash{{\SetFigFont{12}{14.4}{\rmdefault}{\mddefault}{\updefault}$d$}}}}
\put(1078,-557){\makebox(0,0)[lb]{\smash{{\SetFigFont{12}{14.4}{\rmdefault}{\mddefault}{\updefault}$b$}}}}
\put(1264,-531){\makebox(0,0)[lb]{\smash{{\SetFigFont{12}{14.4}{\rmdefault}{\mddefault}{\updefault}$c$}}}}
\put(1592,-1263){\makebox(0,0)[lb]{\smash{{\SetFigFont{12}{14.4}{\rmdefault}{\mddefault}{\updefault}$c$}}}}
\put(1784,-1231){\makebox(0,0)[lb]{\smash{{\SetFigFont{12}{14.4}{\rmdefault}{\mddefault}{\updefault}$d$}}}}
\put(655,-1237){\makebox(0,0)[lb]{\smash{{\SetFigFont{12}{14.4}{\rmdefault}{\mddefault}{\updefault}$b$}}}}
\put(463,-1263){\makebox(0,0)[lb]{\smash{{\SetFigFont{12}{14.4}{\rmdefault}{\mddefault}{\updefault}$a$}}}}
\put(3432,-539){\makebox(0,0)[lb]{\smash{{\SetFigFont{12}{14.4}{\rmdefault}{\mddefault}{\updefault}$c$}}}}
\put(3760,-1271){\makebox(0,0)[lb]{\smash{{\SetFigFont{12}{14.4}{\rmdefault}{\mddefault}{\updefault}$c$}}}}
\put(2624,-1216){\makebox(0,0)[lb]{\smash{{\SetFigFont{12}{14.4}{\rmdefault}{\mddefault}{\updefault}$c$}}}}
\put(2778,-1241){\makebox(0,0)[lb]{\smash{{\SetFigFont{12}{14.4}{\rmdefault}{\mddefault}{\updefault}$d$}}}}
\put(3336,-626){\makebox(0,0)[lb]{\smash{{\SetFigFont{12}{14.4}{\rmdefault}{\mddefault}{\updefault}$a$}}}}
\put(3266,-523){\makebox(0,0)[lb]{\smash{{\SetFigFont{12}{14.4}{\rmdefault}{\mddefault}{\updefault}$b$}}}}
\put(3894,-1318){\makebox(0,0)[lb]{\smash{{\SetFigFont{12}{14.4}{\rmdefault}{\mddefault}{\updefault}$d$}}}}
\put(1091,-1765){\makebox(0,0)[lb]{\smash{{\SetFigFont{12}{14.4}{\rmdefault}{\mddefault}{\updefault}(b)}}}}
\put(2227, 50){\makebox(0,0)[lb]{\smash{{\SetFigFont{12}{14.4}{\rmdefault}{\mddefault}{\updefault}(a)}}}}
\put(3247,-1701){\makebox(0,0)[lb]{\smash{{\SetFigFont{12}{14.4}{\rmdefault}{\mddefault}{\updefault}(c)}}}}
\end{picture}%

%% file: d6.pstex_t
\begin{picture}(0,0)%
\includegraphics{d6.pstex}%
\end{picture}%
\setlength{\unitlength}{3947sp}%
\begingroup\makeatletter\ifx\SetFigFont\undefined%
\gdef\SetFigFont#1#2#3#4#5{%
  \reset@font\fontsize{#1}{#2pt}%
  \fontfamily{#3}\fontseries{#4}\fontshape{#5}%
  \selectfont}%
\fi\endgroup%
\begin{picture}(3299,1971)(1208,-2102)
\put(1572,-1306){\makebox(0,0)[lb]{\smash{{\SetFigFont{12}{14.4}{\rmdefault}{\mddefault}{\updefault}$f$}}}}
\put(1572,-1475){\makebox(0,0)[lb]{\smash{{\SetFigFont{12}{14.4}{\rmdefault}{\mddefault}{\updefault}$c$}}}}
\put(1805,-1299){\makebox(0,0)[lb]{\smash{{\SetFigFont{12}{14.4}{\rmdefault}{\mddefault}{\updefault}$e$}}}}
\put(1819,-1504){\makebox(0,0)[lb]{\smash{{\SetFigFont{12}{14.4}{\rmdefault}{\mddefault}{\updefault}$d$}}}}
\put(1589,-410){\makebox(0,0)[lb]{\smash{{\SetFigFont{12}{14.4}{\rmdefault}{\mddefault}{\updefault}$j$}}}}
\put(1560,-565){\makebox(0,0)[lb]{\smash{{\SetFigFont{12}{14.4}{\rmdefault}{\mddefault}{\updefault}$a$}}}}
\put(1814,-424){\makebox(0,0)[lb]{\smash{{\SetFigFont{12}{14.4}{\rmdefault}{\mddefault}{\updefault}$e$}}}}
\put(1793,-600){\makebox(0,0)[lb]{\smash{{\SetFigFont{12}{14.4}{\rmdefault}{\mddefault}{\updefault}$c$}}}}
\put(2442,-1736){\makebox(0,0)[lb]{\smash{{\SetFigFont{12}{14.4}{\rmdefault}{\mddefault}{\updefault}$d$}}}}
\put(2421,-1927){\makebox(0,0)[lb]{\smash{{\SetFigFont{12}{14.4}{\rmdefault}{\mddefault}{\updefault}$e$}}}}
\put(2619,-1821){\makebox(0,0)[lb]{\smash{{\SetFigFont{12}{14.4}{\rmdefault}{\mddefault}{\updefault}$g$}}}}
\put(2880,-911){\makebox(0,0)[lb]{\smash{{\SetFigFont{12}{14.4}{\rmdefault}{\mddefault}{\updefault}$c$}}}}
\put(3084,-742){\makebox(0,0)[lb]{\smash{{\SetFigFont{12}{14.4}{\rmdefault}{\mddefault}{\updefault}$b$}}}}
\put(3141,-953){\makebox(0,0)[lb]{\smash{{\SetFigFont{12}{14.4}{\rmdefault}{\mddefault}{\updefault}$d$}}}}
\put(4016,-431){\makebox(0,0)[lb]{\smash{{\SetFigFont{12}{14.4}{\rmdefault}{\mddefault}{\updefault}$b$}}}}
\put(4009,-650){\makebox(0,0)[lb]{\smash{{\SetFigFont{12}{14.4}{\rmdefault}{\mddefault}{\updefault}$h$}}}}
\put(4185,-516){\makebox(0,0)[lb]{\smash{{\SetFigFont{12}{14.4}{\rmdefault}{\mddefault}{\updefault}$i$}}}}
\end{picture}%

%% file: 56proof1.pstex_t
\begin{picture}(0,0)%
\includegraphics{56proof1.pstex}%
\end{picture}%
\setlength{\unitlength}{3947sp}%
\begingroup\makeatletter\ifx\SetFigFont\undefined%
\gdef\SetFigFont#1#2#3#4#5{%
  \reset@font\fontsize{#1}{#2pt}%
  \fontfamily{#3}\fontseries{#4}\fontshape{#5}%
  \selectfont}%
\fi\endgroup%
\begin{picture}(4053,2124)(431,-1375)
\put(752,-25){\makebox(0,0)[lb]{\smash{{\SetFigFont{12}{14.4}{\rmdefault}{\mddefault}{\updefault}$x$}}}}
\put(1466,-1021){\makebox(0,0)[lb]{\smash{{\SetFigFont{12}{14.4}{\rmdefault}{\mddefault}{\updefault}$J$}}}}
\put(2527,-1293){\makebox(0,0)[lb]{\smash{{\SetFigFont{12}{14.4}{\rmdefault}{\mddefault}{\updefault}$\Gamma(J)$}}}}
\put(4161,-861){\makebox(0,0)[lb]{\smash{{\SetFigFont{12}{14.4}{\rmdefault}{\mddefault}{\updefault}$y$}}}}
\end{picture}%

%% file: 56proof2.pstex_t
\begin{picture}(0,0)%
\includegraphics{56proof2.pstex}%
\end{picture}%
\setlength{\unitlength}{3947sp}%
\begingroup\makeatletter\ifx\SetFigFont\undefined%
\gdef\SetFigFont#1#2#3#4#5{%
  \reset@font\fontsize{#1}{#2pt}%
  \fontfamily{#3}\fontseries{#4}\fontshape{#5}%
  \selectfont}%
\fi\endgroup%
\begin{picture}(4253,2124)(431,-1375)
\put(752,-25){\makebox(0,0)[lb]{\smash{{\SetFigFont{12}{14.4}{\rmdefault}{\mddefault}{\updefault}$x$}}}}
\put(4161,-861){\makebox(0,0)[lb]{\smash{{\SetFigFont{12}{14.4}{\rmdefault}{\mddefault}{\updefault}$y$}}}}
\put(2548,-1291){\makebox(0,0)[lb]{\smash{{\SetFigFont{12}{14.4}{\rmdefault}{\mddefault}{\updefault}$J$}}}}
\put(3927,-1188){\makebox(0,0)[lb]{\smash{{\SetFigFont{12}{14.4}{\rmdefault}{\mddefault}{\updefault}$C_{l}$}}}}
\put(1213,-1079){\makebox(0,0)[lb]{\smash{{\SetFigFont{12}{14.4}{\rmdefault}{\mddefault}{\updefault}$C_{j}$}}}}
\put(1098,-136){\makebox(0,0)[lb]{\smash{{\SetFigFont{12}{14.4}{\rmdefault}{\mddefault}{\updefault}$i$}}}}
\put(3997,-514){\makebox(0,0)[lb]{\smash{{\SetFigFont{12}{14.4}{\rmdefault}{\mddefault}{\updefault}$k$}}}}
\end{picture}%

%% file: levels.pstex_t
\begin{picture}(0,0)%
\includegraphics{levels.pstex}%
\end{picture}%
\setlength{\unitlength}{3947sp}%
\begingroup\makeatletter\ifx\SetFigFont\undefined%
\gdef\SetFigFont#1#2#3#4#5{%
  \reset@font\fontsize{#1}{#2pt}%
  \fontfamily{#3}\fontseries{#4}\fontshape{#5}%
  \selectfont}%
\fi\endgroup%
\begin{picture}(4241,3970)(801,-3656)
\put(2894,-2580){\makebox(0,0)[lb]{\smash{{\SetFigFont{12}{14.4}{\rmdefault}{\mddefault}{\updefault}level $l+1$}}}}
\put(4598,-1837){\makebox(0,0)[lb]{\smash{{\SetFigFont{12}{14.4}{\rmdefault}{\mddefault}{\updefault}$u$}}}}
\put(3668,-1878){\makebox(0,0)[lb]{\smash{{\SetFigFont{12}{14.4}{\rmdefault}{\mddefault}{\updefault}$v$}}}}
\put(3369,-3004){\makebox(0,0)[lb]{\smash{{\SetFigFont{12}{14.4}{\rmdefault}{\mddefault}{\updefault}level $l$}}}}
\put(4299,-2663){\makebox(0,0)[lb]{\smash{{\SetFigFont{12}{14.4}{\rmdefault}{\mddefault}{\updefault}level $l-1$}}}}
\end{picture}%

%% file: alg_g.pstex_t
\begin{picture}(0,0)%
\includegraphics{alg_g.pstex}%
\end{picture}%
\setlength{\unitlength}{3947sp}%
\begingroup\makeatletter\ifx\SetFigFont\undefined%
\gdef\SetFigFont#1#2#3#4#5{%
  \reset@font\fontsize{#1}{#2pt}%
  \fontfamily{#3}\fontseries{#4}\fontshape{#5}%
  \selectfont}%
\fi\endgroup%
\begin{picture}(2631,2301)(508,-2786)
\end{picture}%

%% file: alg_d.pstex_t
\begin{picture}(0,0)%
\includegraphics{alg_d.pstex}%
\end{picture}%
\setlength{\unitlength}{3947sp}%
\begingroup\makeatletter\ifx\SetFigFont\undefined%
\gdef\SetFigFont#1#2#3#4#5{%
  \reset@font\fontsize{#1}{#2pt}%
  \fontfamily{#3}\fontseries{#4}\fontshape{#5}%
  \selectfont}%
\fi\endgroup%
\begin{picture}(5102,3878)(602,-3575)
\end{picture}%

%% file: table1.pstex_t
\begin{picture}(0,0)%
\includegraphics{table1.pstex}%
\end{picture}%
\setlength{\unitlength}{3947sp}%
\begingroup\makeatletter\ifx\SetFigFont\undefined%
\gdef\SetFigFont#1#2#3#4#5{%
  \reset@font\fontsize{#1}{#2pt}%
  \fontfamily{#3}\fontseries{#4}\fontshape{#5}%
  \selectfont}%
\fi\endgroup%
\begin{picture}(4075,4013)(1376,-4324)
\end{picture}%

%% file: drawing1.pstex_t
\begin{picture}(0,0)%
\includegraphics{drawing1.pstex}%
\end{picture}%
\setlength{\unitlength}{3947sp}%
\begingroup\makeatletter\ifx\SetFigFont\undefined%
\gdef\SetFigFont#1#2#3#4#5{%
  \reset@font\fontsize{#1}{#2pt}%
  \fontfamily{#3}\fontseries{#4}\fontshape{#5}%
  \selectfont}%
\fi\endgroup%
\begin{picture}(6500,6750)(2476,-7786)
\end{picture}%

%% file: plane.bbl
\begin{thebibliography}{10}

\bibitem{arnborg1}
Stefan Arnborg.
\newblock Efficient algorithms for combinatorial problems on graphs with
  bounded decomposability.
\newblock {\em BIT}, 25(1):2--23, 1985.

\bibitem{arnborg2}
Stefan Arnborg and Andrzej Proskurowski.
\newblock Linear time algorithms for np-hard problems restricted to partial
  k-trees.
\newblock {\em Discrete Appl. Math.}, 23(1):11--24, 1989.

\bibitem{baker}
Brenda~S. Baker.
\newblock Approximation algorithms for np-complete problems on planar graphs.
\newblock {\em J. ACM}, 41(1):153--180, 1994.

\bibitem{Bienstock}
Daniel Bienstock and Clyde~L. Monma.
\newblock On the complexity of embedding planar graphs to minimize certain
  distance measures.
\newblock {\em Algorithmica}, 5:93--109, 1990.

\bibitem{bodlaender96}
Hans~L. Bodlaender.
\newblock A linear-time algorithm for finding tree-decompositions of small
  treewidth.
\newblock {\em SIAM J. Comput.}, 25(6):1305--1317, 1996.

\bibitem{bodlaender97}
Hans~L. Bodlaender.
\newblock Treewidth: Algorithmic techniques and results.
\newblock {\em Proceedings 22nd International Symposium on Mathmatical
  Foundations of Computer Science, MFCS'97}, 1295:29--36, 1997.

\bibitem{chiba}
Norishige Chiba, Takao Nishizeki, and Nobuji Saito.
\newblock An approximation algorithm for the maximum independent set problem on
  planar graphs.
\newblock {\em SIAM J. Comput. 11}, 4:663--675, 1982.

\bibitem{dmw}
Vida Dujmovi\'c, Pat Mworin, and David~R. Wood.
\newblock Pathwidth and three-dimensional straight line grid drawings of
  graphs.
\newblock {\em Lecture Notes in Comp. Science, Proc. 10th Int'l Symp. Graph
  Drawing}, pages 42--53, 2002.

\bibitem{eades88}
Peter Eades and Roberto Tamassia.
\newblock Algorithms for drawing graphs: An annotated bibliography, 1988.

\bibitem{garey}
Michael.~R. Garey and David~S. Johnson.
\newblock {\em Computers and Intractability, A guide to the Theory of
  NP-Completeness}.
\newblock W.H.Freeman and Co,, 1979.

\bibitem{hadlock}
F.~Hadlock.
\newblock Finding a maximum cut of a planar graph in polynomial time.
\newblock {\em SIAM J. Comput}, 4:221--225, 1975.

\bibitem{harary}
Frank Harary.
\newblock {\em Graph Theory}.
\newblock Addison-Wesley, 1971.

\bibitem{hopcroftb}
John~E. Hopcroft, Rajeev Motwani, and Jeffrey~D. Ullman.
\newblock {\em Introduction to Automata Theory, Languages, and Computation, 2nd
  Edition}.
\newblock Addison-Wesley, 2000.

\bibitem{hopcroft74}
John~E. Hopcroft and Robert Tarjan.
\newblock Efficient planarity testing.
\newblock {\em J. ACM}, 21(4):549--568, 1974.

\bibitem{junger}
Michael J\"unger, Petra Mutzel, Thomas Odenthal, and Mark Scharbrodt.
\newblock The thickness of a minor-excluded class of graphs.
\newblock {\em Discrete Math.}, 182(1-3):169--176, 1998.

\bibitem{kloks}
Ton Kloks.
\newblock {\em Treewith Computations and Approximations}, volume 842.
\newblock Springer-Verlag, Berlin/NY, 1994.

\bibitem{liebers}
Annegret Liebers.
\newblock Planarizing graphs - a survey and annotated bibliography.
\newblock {\em Journal of Graph Algorithms and Applications}, 5(1):1--74, 2001.

\bibitem{lipton}
Richard~J. Lipton and Robert~E. Tarjan.
\newblock Applications of a planar separator theorem.
\newblock {\em SIAM J. Comput. 9}, 3:615--627, 1980.

\bibitem{purchase}
Helen~C. Purchase.
\newblock The effects of graph layout.
\newblock {\em OZCHI '98: Proceedings of the Australasian Conference on
  Computer Human Interaction}, pages 80--86, 1998.

\bibitem{roberts96}
Neil Robertson, Daniel~P. Sanders, Paul~D. Seymour, and Robin Thomas.
\newblock Efficiently four-coloring planar graphs.
\newblock In {\em STOC '96: Proceedings of the twenty-eighth annual ACM
  symposium on Theory of computing}, pages 571--575, New York, NY, USA, 1996.
  ACM Press.

\bibitem{minor1}
Neil Robertson and Paul~D. Seymour.
\newblock Graph minors. i. excluding a forest.
\newblock {\em Journal of Combinatorial Theory Series B}, 35:39--61, 1983.

\bibitem{minor2}
Neil Robertson and Paul~D. Seymour.
\newblock Graph minors. ii. algorithmic aspects of tree-width.
\newblock {\em Journal of Algorithms}, 7:309--322, 1986.

\bibitem{west}
Douglas~B. West.
\newblock {\em Graph Theory}.
\newblock Prentice Hall, Upper Saddle River, N.J., 2001.

\end{thebibliography}
